 \documentclass[final,authoryear,3p,times]{elsarticle}

\usepackage{amssymb}
\usepackage{rotating}
\usepackage{multirow,color}
\usepackage{ulem}
\usepackage{mathrsfs}
\usepackage{url}

\journal{Journal of International Financial Markets, Institutions \& Money}

\begin{document}

\begin{frontmatter}



\title{Time-dependent lead-lag relationship\\ between the onshore and offshore Renminbi exchange rates
    \let\thefootnote\relax\footnotetext{We thank participants at the $4^{\rm{th}}$ International Symposium in Computational Economics and Finance, Paris (April 2016). We are grateful to Fredj Jawadi, Shu-Heng Chen, Imran Shah, Estrella Gomez-Herrera, Yuyi Li, Yi-Heng Tseng and Yanyi Wang for insightful discussions of this work. Financial supports from the National Natural Science Foundation of China Grants (71501072), the China Postdoctoral Science Foundation Grants (2015M570342) and the Fundamental Research Funds for the Central Universities are acknowledged.}
    }
\author[BS,PRS]{Hai-Chuan Xu}
\author[BS,SS]{Wei-Xing Zhou \corref{cor}}
\ead{wxzhou@ecust.edu.cn}
\author[ETH,SFI]{Didier Sornette}
\cortext[cor]{Corresponding author.}
\ead{dsornette@ethz.ch} %

\address[BS]{Department of Finance, East China University of Science and Technology, Shanghai 200237, China}
\address[PRS]{Postdoctoral Research Station, East China University of Science and Technology, Shanghai 200237, China}
\address[SS]{Department of Mathematics, East China University of Science and Technology, Shanghai 200237, China}
\address[ETH]{Department of Management, Technology and Economics, ETH Zurich, Zurich, Switzerland}
\address[SFI]{Swiss Finance Institute, c/o University of Geneva, 40 blvd. Du Pont d¡¯Arve, CH 1211 Geneva 4, Switzerland}

\begin{abstract}
  We employ the thermal optimal path method to explore both the long-term and short-term interaction patterns between the onshore CNY and offshore CNH exchange rates (2012-2015). For the daily data, the CNY and CNH exchange rates show a weak alternate lead-lag structure in most of the time periods. When CNY and CNH display a large disparity, the lead-lag relationship is uncertain and depends on the prevailing market factors. The minute-scale interaction pattern between the CNY and CNH exchange rates change over time according to different market situations. We find that US dollar appreciation is associated with a lead-lag relationship running from offshore to onshore, while a (contrarian) Renminbi appreciation is associated with a lead-lag relationship running from onshore to offshore. These results are robust with respect to different sub-sample analyses and variations of the key smoothing parameter of the TOP method.
\end{abstract}

\begin{keyword}
Renminbi exchange rates; Onshore and offshore markets; Lead-lag structure; Thermal optimal path

 \medskip
\noindent \textit{JEL classification}: C14, F31, G15.
\end{keyword}

\end{frontmatter}

\newpage

\section{Introduction}
\label{S1:Introduction}

The Renminbi (RMB) has witnessed an increasing influence worldwide, paralleling the ascendency of China
as the second economic powerhouse. As of February 2015, Chinese RMB has consolidated
its position as the second most used currency for documentary credit trade transactions\footnote{\url{http://www.forbes.com/sites/rogeraitken/2015/02/28/chinese-rmb-consolidates-second-most-used-currency-ranking-for-dc-trade-transactions/#12a5a7382d8c}}
and fourth most-used payments currency globally\footnote{\url{http://www.bloomberg.com/news/articles/2015-10-06/yuan-overtakes-yen-as-world-s-fourth-most-used-payments-currency}}, since October 2015, according to the SWIFT RMB Tracker.
According to the ``Triennial Central Bank Survey on the foreign exchange market'' released by the \cite{BIS-2013-Basel}, in 2013, Renminbi ranked as the world's 9th most traded currency with an average daily turnover of 119.56 billion dollars.
As of December 2015, in terms of currency trading, the RMB remains at the 9th rank\footnote{\url{http://bytefreaks.net/other/most-traded-currencies-by-value-december-201}}.


In this paper, we focus on one currency (i.e. RMB) traded in two separate markets. The RMB transacted in onshore market has the trading symbol CNY, while the RMB transacted in offshore market has the trading symbol CNH. Therefore, both CNY rate and CNH rate are the pricing of RMB, but in different markets. According to the law of one price, if there are no capital constraints and other market frictions, the rates would be expected to be the same. However, in reality, the CNY market is constrained by the central bank's intervention and limited by a daily trading band. The band against the US dollar was increased from 0.3\% to 0.5\% in 2007 and to 1\% in 2010, then widened again to 2\% in March 2014. On the contrary, the CNH market is a free market. This will make it possible to have two distinct exchange rates for the same currency RMB. Furthermore, China's capital account is one of the least open in the world according to the Chinn-Ito index, a widely used indicator for capital account openness \citep{Chinn-Ito-2006-JDE}. The control of cross-border capital flow in mainland China causes that the pricing gaps cannot be easily arbitraged away.

Due to the limit of arbitrage, it is a normal state that there exists pricing gap between CNY rate and CNH rate. For example, on 26 January 2014 the CNH/USD exchange rate (6.0414) was 74 pips (price interest points) below the CNY/USD rate (6.0488), while by 5 February the gap was widened to 326 pips (6.0274 for CNH vs. 6.0600 for CNY). However, only two weeks later on 21 February the differential declined to 15 pips. This shows that the deviation between CNY and CNH is time-varying, even the CNY and CNH exchange rates usually track each other quite well. \cite{Craig-Hua-Ng-Yuen-2013-IMF} claim that the shifts in global market sentiment and capital controls can explain much of the divergence in the CNY and CNH exchange rates. According to extended GARCH models, \cite{Funke-Shu-Cheng-Eraslan-2015-JIMF} find that the liquidity diversity between the two markets plays an important role in explaining the CNY-CNH pricing differential. For Renminbi derivatives, \cite{Wang-2015-JBF} proposes a theoretical model and finds that the conversion restrictions in the spot market rather than the changes in credit risk or liquidity constraint cause the covered interest parity (CIP) deviations observed in the onshore Renminbi forward market. \cite{Li-Hui-Chung-2012-HKIMR} find that model parameter uncertainty can explain price disparities between the Renminbi onshore deliverable forward and offshore non-deliverable forward exchange rates. From the perspective of microstructure, \cite{Zhang-Chau-Zhang-2013-IRFA} argue that the order flow influences both the long-term level and short-term fluctuations of the Renminbi exchange rate.

Besides the investigation for the reason of CNY-CNH price disparities, the relationship between the CNY and CNH exchange rates has also been studied. With a general belief that, under tight capital controls, investors may choose to alternatively trade in the offshore market, the onshore price discovery should be a presumed function of the offshore exchange rate. Indeed, \cite{Cheung-Rime-2014-JIMF} use order flow data and VECM regression to study the CNH exchange rate dynamics and its links with the CNY exchange rate. They find that the offshore CNH exchange rate has an increasing impact on the onshore CNY rate, and has predictive power for the official Renminbi central parity rate. Similarly, \cite{Shu-He-Cheng-2015-CER} indicate that the CNH rate tends to have greater impacts on regional currencies in normal market conditions, but the CNY rate has greater impacts during periods of market stress. However, \cite{Leung-Fu-2014-HKIMR} study the interactions between the Renminbi forward market in Mainland China and Hong Kong and find that the onshore to offshore impact is larger than that in the opposite direction in most cases. Differently, \cite{Maziad-Kang-2012-IMF} employ a bivariate GARCH model to understand the inter-linkages between CNY and CNH markets and claim that, although developments in the onshore spot market exert an influence on the offshore spot market, offshore forward rates have a predictive impact on onshore forward rates. In addition, \cite{Ding-Tse-Williams-2014-JFutM} believe that price discovery is absent between the onshore and offshore spot markets, but present between onshore spot and offshore non-deliverable forward rates.

Motivated by the controversy of the interaction relationship between the CNH and CNY rates, this paper intents to test the time-varying lead-lag interaction structure between these two exchange rates. Further, considering that the onshore and offshore exchange rates exhibit both short-term and long-term interactions, we investigate daily interaction pattern and minute-scale interaction pattern for the CNY and CNH exchange rates.

Instead of the conventional Granger causality method, we employ the thermal optimal path method to capture the dynamic lead-lag structure between the CNY and CNH exchange rates. According to different weight designs, such new approaches can be divided into two categories: thermal optimal path (TOP) with time-forward weights \citep{Sornette-Zhou-2005-QF,Zhou-Sornette-2006-JMe} and thermal optimal path with time-reversed symmetric weights (TOPS) \citep{Meng-Xu-Zhou-Sornette-2017-QF}. These methods have been applied to explore the lead-lag relationships in many areas, such as inflation/GDP, stock market/GDP, stock market returns/bond interests, house price/monetary policy, stock index/stock index futures and so on \citep{Zhou-Sornette-2007-PA,Sornette-Zhou-2005-QF,Guo-Zhou-Cheng-Sornette-2011-PLoS1,Guo-Zhou-Cheng-2012-cnJMSC,Meng-Xu-Zhou-Sornette-2017-QF,Gong-Ji-Su-Li-Ren-2016-PA}. One of the main advantages of the TOP/TOPS methods is that they are able to identify the time-varying lead-lag structure between two time series, in contrast to the fixed period forward causality detection in the literature. Given the time-varying properties, this non-parametric method does not require the stationarity of time series. This feature means that one do not need to differentiate the non stationary series, thus preserving all the information of the original time series.

The paper is organized as follows. Section~\ref{S1:Data:method} describes the data we will investigate and summarises the TOP method we will use. Section~\ref{S1:Interaction:patterns} analyses the time-dependent lead-lag structure between CNY rate and CNH rate at the daily frequency and the minute-scale frequency respectively. For robustness, we also provide a sub-sample analysis, corresponding to episodes of price increasing, fall and oscillation. Section~\ref{S1:Alternative:specifications} does an alternative analysis with different optimization parameters. Section~\ref{S1:Conclusions} concludes.

\section{Data and methods}
\label{S1:Data:method}

\subsection{Data}

The daily exchange rates for the long-term CNY-CNH interaction analysis span from 30 April 2012 to 22 October 2015. The minute-scale exchange rates for the short-term CNY-CNH interaction analysis cover from 11 September 2015 to 23 October 2015. The foreign exchange settlement and sale business run from 9:30 to 16:30 in the onshore market\footnote{The People's Bank of China announced that the closing time of foreign exchange trading would be extended to 23:30 since 4 January 2016 (see http://www.pbc.gov.cn/goutongjiaoliu/113456/113469/2994220/index.html). We also tested some datasets after 4 January 2016 in the following and the results in this work are not influenced by the extension of trading time.} and open 24 hours in the offshore market. For daily datasets, the CNY rates we used are the closing prices at 16:30 every day. In order to guarantee the synchronization of datasets, we choose the rates at the same time point 16:30 as the CNH rates. For the minute-scale datasets, considering that the CNY market opens from 9:30 to 16:30, it is natural that we choose the CNH rates in the same time period. The minute-scale datasets cover all market states of bull, bear and stable. All data are from the ``WIND'' database.

\subsection{Thermal optimal path method}

The TOP method was proposed by \cite{Sornette-Zhou-2005-QF}. When the relationship between two time series $\{X_{t_1}: t_1 = 1,\ldots,N\}$ and $\{Y_{t_2}: t_2 = 1,\ldots,N\}$ is more complex than a simple constant lead-lag form $Y_t = a X_{t-k} + \eta_t$, the correspondence between the two time series is no longer obvious. An intuitive idea is to search for the dependence relationship in the form of a one-to-one mapping $t_2 = \phi(t_1)$ between the time series $X_{t_1}$ and $Y_{t_2}$ such that the two time series are closest in some sense. We first form a distance matrix that allows us to compare all values of $X_{t_1}$ with all values of $Y_{t_2}$
along the two time axes $t_1$ and $t_2$. The elements of the distance matrix are defined as
\begin{equation}\label{Eq:Distance}
  \varepsilon(t_1, t_2)= \varepsilon(t_1, t_2)_{-} = |X_{t_1} - Y_{t_2}|.
\end{equation}
The distance matrix further defines the mapping $t_1\longrightarrow t_2 = \phi(t_1)$ as follows,
\begin{equation}\label{Eq:Mapping}
  \phi(t_1) = \min_{t_2}{\varepsilon(t_1, t_2)}.
\end{equation}
This mapping makes the distance minimum over all possible $t_2$ for a fixed $t_1$. Considering the simplest example like $Y_{t} = X_{t-k}$ with a positive constant $k$, we have $\varepsilon(t_1, t_2) =0$ for $t_2 = t_1 + k$, which is parallel to the main diagonal of the distance matrix. This line equivalently defines the mapping $t_2 = \phi(t_1)=t_1 + k$. When $k=0$, the mapping is exactly the main diagonal of the distance matrix.

We should stress that, as \cite{Zhou-Sornette-2006-JMe,Zhou-Sornette-2007-PA} pointed out, the definition of distance matrix (\ref{Eq:Distance}) is only proper when the two time series $X$ and $Y$ display the co-monotonic relationship. When two time series $X$ and $Y$ display the anti-monotonic relationship, i.e., they tend to take opposite signs, the elements of the distance matrix should be defined as $\varepsilon(t_1, t_2)_{+} = |X_{t_1} + Y_{t_2}|$. The $+$ sign ensures a correct search of synchronization between two anti-correlated time series. One can think the example $Y (t) = - X(t - k)$. More generally, $X$ and $Y$ may exhibit more complicated lead-lag correlation relationships, positive correlation over some time intervals and negative correlation at other times. In order to address all possible situations, \cite{Zhou-Sornette-2006-JMe,Zhou-Sornette-2007-PA} propose to use the mixed distance expressed as $\varepsilon(t_1, t_2)_{\pm} = \min[\varepsilon(t_1, t_2)_{-}, \varepsilon(t_1, t_2)_{+}]$. In current paper, due to the fact that both CNY and CNH are the pricing of RMB and that they always co-move and have significantly positive correlation, the distance defined by (\ref{Eq:Distance}) is a naturally selection.

Note that the minimization process (\ref{Eq:Mapping}) analyzes $t_1$ each time independently of others. This causes two undesirable features. First, in the presence of noise, with large probability, there will be many $t_1$ such that the increments $\phi(t_1 +1) - \phi(t_1)$ have big jumps. However, a continuous lag function of time is more expected and large jumps in the function $\phi$ are not reasonable. Second, a one-to-one mapping might be absent, that is, a given $t_1$ would be associated with several $t_2$. To address these two problems, \cite{Sornette-Zhou-2005-QF} replace the mapping in Eq.~(\ref{Eq:Mapping}) determined by a local minimization by a mapping obtained by the following global minimization
\begin{equation}\label{Eq:Global:Min}
  \min_{\{\phi(t_1),t_1=1,2,...,N\}}\sum_{t_1=1}^{N}|X_{t_1}-Y_{\phi(t_1)}|,
\end{equation}
with a smoothing constraint
\begin{equation}\label{Eq:Smooth}
  0\leqslant \phi(t_1 + 1) - \phi(t_1) \leqslant 1.
\end{equation}
In the continuous-time limit of this condition, the mapping $t_1 \rightarrow t_2=\phi(t_1)$ can be viewed as continuous.

The existence of the continuity constraint transforms the problem into a global optimization problem. This problem actually has a long research history in statistical physics \citep{HalpinHealy-Zhang-1995-PR}, called ``random directed polymer at zero temperature''. Indeed, the distance matrix (\ref{Eq:Distance}) can be seen as the energy landscape in the plane $(t_1,t_2)$, The continuity constraint means that the mapping defines a path or ``polymer'' of equation $t_2 = \phi(t_1)$ with a ``line tension'' to prevent discontinuous. The condition that $\phi(t_1)$ is non-decreasing means that the polymer should be directed. Then we can translate the global minimization problem (\ref{Eq:Global:Min}) into searching for the polymer configuration with minimum energy. Next, we will describe the implementation of the search for the optimal lag path between two time series.

\begin{figure}
  \centering
  \includegraphics[width=0.5\linewidth]{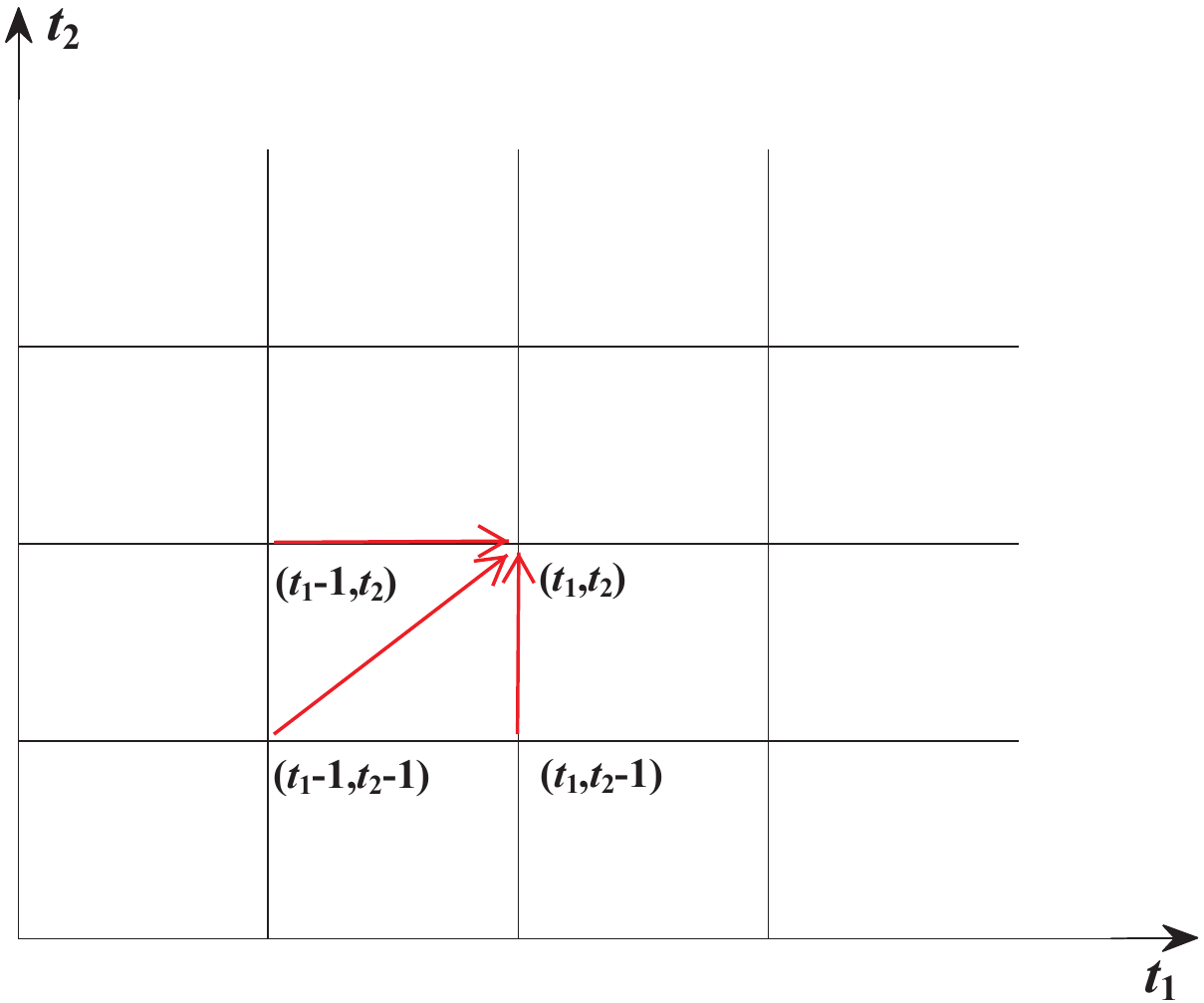}\\
  \caption{A simple schematic representation of the thermal optimal path method with time-forward weight in the lattice ($t_1,t_2$). The $\rightarrow$ denotes that the weight process is recursively from three directions.}\label{Fig:TOP:Lattice}
\end{figure}

To solve the global optimization problem (\ref{Eq:Global:Min}) in polynomial time,
the transfer matrix method in the rotated coordinate system has been suggested \citep{Derrida-Vannimenus-Pomeau-1978-JPC,Derrida-Vannimenus-1983-PRB}. A given node $(t_1, t_2)$ carries the ``potential energy'' or distance $\varepsilon(t_1, t_2)$. Let us start with the time-forward direction. Under the continuity constraint (\ref{Eq:Smooth}), the optimal path can either go along the $t_1$-axis by one step from $(t_1, t_2)$ to $(t_1 +1, t_2)$, along the $t_2$-axis by one step from $(t_1,t_2)$ to $(t_1, t_2 +1)$ or along the diagonal from $(t_1, t_2)$ to $(t_1+1, t_2 +1)$ (refer to the red arrow in Fig.~\ref{Fig:TOP:Lattice}). Let $E(t_1,t_2)$ be the cumulative energy of the optimal path starting from some origin $(t_{1,0}, t_{2,0})$ and ending at $(t_1, t_2)$. Then, the following recursive relation can be established:
\begin{equation}\label{Eq:Recursive:E}
  E(t_1,t_2) = \varepsilon(t_1,t_2) + \min[E(t_1-1,t_2),E(t_1,t_2-1),E(t_1-1,t_2-1)].
\end{equation}
This means that the minimum energy path reaching $(t_1, t_2)$ is nothing but a recursive extension of the minimum energy path reaching one of these three previous points. The global minimization process is completely determined once the starting point and the ending point of the path are defined. The minimum energy path over all possible starting and ending points is the solution of the global optimization (\ref{Eq:Global:Min}). The solution $t_2 = \phi(t_1)$ reflects the time-dependent lag structure between the two time series.

In reality, financial time series are noisy so that their corresponding energy landscape may contain spurious patterns and lead to incorrect lead-lag relationships. To obtain a robust lead-lag path, \cite{Sornette-Zhou-2005-QF} suggest to introduce ``thermal excitations'' around the path such that path configurations with slightly larger global energies are allowed to contribute to the conformation of the optimal path with probability decreasing with their energy. The ``temperature'' $T$ they specified quantifies how much deviations from the minimum energy are allowed. Increasing $T$ allows to sample more and more paths around the minimum energy path, thus allows one to obtain an average ``optimal thermal path'' over a larger and larger number of path conformations, leading to more robust estimates of the lead-lag structure between the two time series. Of course, for too large temperatures, the energy landscape or distance matrix becomes irrelevant and one looses all information in the lead-lag relationship between the two time series.

The thermal optimal path method (TOP) starts with the allocation of weights to each node $(t_1, t_2)$. The weight process of the TOP method is recursively from left to right (time-forward direction). As illustrated in Fig.~\ref{Fig:TOP:Lattice}, to arrive at $(t_1, t_2)$, the path can either from $(t_1,t_2-1)$, $(t_1-1,t_2)$ or $(t_1-1,t_2-1)$. The recursive equation on the Boltzmann weight factor of
each node (in the presence of the finite temperature that enables sampling of paths)  is thus
\begin{equation}\label{Eq:Recursive:G}
  G(t_1,t_2) = [G(t_1,t_2-1)+G(t_1-1,t_2)+G(t_1-1,t_2-1)]e^{-\varepsilon(t_1,t_2)/T},
\end{equation}
where $\varepsilon(t_1,t_2)$ is the local energy determined by the distance matrix element (\ref{Eq:Distance}) at node $(t_1, t_2)$.

It is convenient to rotate the coordinate system such that $(t_1, t_2)\rightarrow (x, \tau)$ where $\tau$ is the main diagonal direction of the $(t_1, t_2)$ system and $x$ is orthogonal to $\tau$. The transformation from the coordinates ($t_1,t_2$) to ($x,\tau$) is given by
\begin{equation}\label{Eq:Transfer}
  \left\{
  \begin{array}{l}
    x = t_2 - t_1, \\
    \tau = t_2 + t_1 - 2.
  \end{array}
  \right.
\end{equation}
If $t_1$ and $t_2$ both start with 1 ($t_1=1,t_2=1$), then the origin of the rotated system is ($x=0,\tau=0$). The first coordinate $x$ quantifies the lead or lag path between two time series. As mentioned above, to obtain a robust lead-lag relationship, we resort to an average optimal thermal path, which is given by the average position over all possible paths weighted by their corresponding
Boltzmann factor:
\begin{equation}\label{Eq:TOP:Average}
  \langle x(\tau) \rangle = \sum_{x}x\mathscr{G}(x,\tau)/\mathscr{G}(\tau),
\end{equation}
where $\mathscr{G}(\tau)=\sum_{x}\mathscr{G}(x,\tau)$, which is the sum of Boltzmann weight factors at $\tau$ over all three paths
ending at a given $(x, \tau)$. And $\mathscr{G}(x,\tau)$ is the transformed function of (\ref{Eq:Recursive:G}), which has the following form
\begin{equation}\label{Eq:Recursive:G:Transformed}
  \mathscr{G}(x,\tau+2) = [\mathscr{G}(x-1,\tau+1) + \mathscr{G}(x+1,\tau+1) + \mathscr{G}(x,\tau)]e^{-\epsilon(x,\tau+2)/T},
\end{equation}
where $\epsilon(x,\tau+2)$ is the coordinate transformed function of $\varepsilon(t_1,t_2)$. This implies that, for larger distances between the two time series at node $(x,\tau)$, the Boltzmann weight factor for the lead-lag path $x$ is smaller, which is an automatic mechanism for searching for the average optimal path. Hence, Eq.~\ref{Eq:TOP:Average} indeed defines $\langle x(\tau) \rangle$ as the thermal average of the local time lag over all lead-lag configurations weighted by the Boltzmann weight factors. This justifies to call it the ``thermal averaged path''.

As we discussed above, once we fix a starting and an ending point, we can determine the thermal optimal path $\langle x(\tau) \rangle$ under a certain temperature $T$. We can also define a cost ``energy'' $e_T$ for this path using the thermal average of the ``energy'' $\epsilon(x,\tau)$ (i.e. ``distance'' (\ref{Eq:Distance}) between the two time series):
\begin{equation}\label{Eq:Energy}
  e_T = \frac{1}{2N-|x_0|-|x_N|-1}\sum_{\tau=|x_0|}^{2N-|x_N|-1}\sum_x \epsilon(x,\tau)\mathscr{G}(x,\tau)/\mathscr{G}(\tau),
\end{equation}
where $N$ is the length of the time series and $|x_0|$ and $|x_N|$ are the absolute values of the lead-lag relationship at the starting and ending points.

From the description of the TOP method, we find that it does not address the fundamental philosophical and epistemological question of the real causality links between $X$ and $Y$. In fact, even the Granger causality concept does not address this question. The Granger causality takes a pragmatic approach based on predictability: if the knowledge of $X(t)$ and of its past values improves the prediction of $Y(t + s)$ for some $s > 0$, then it is said that $X$ Granger causes $Y$ \citep{Ashley-Granger-Schmalensee-1980-Em,Geweke-1984}. Our approach is similar in that it does not address the question of the existence of a genuine causality but attempts to detect a dependence structure between two time series at non-zero lags.

\subsection{Consistency test}

If $\langle x(t) \rangle$ is the lead-lag path identified by the TOP method, synchronising the two time series using the time varying $\langle x(t) \rangle$ should lead to a statistically significant correlation, or at least lead to statistically higher correlation than for the non-synchronising case. More specifically, if $\langle x(t) \rangle$ is indeed interpreted as the lead-lag time between the two time series, this implies that $X_{t-\langle x(t)\rangle}$ and $Y_t$ should be synchronised and should exhibit a strong linear dependence. This forms
the basis of the consistency test for $\langle x(t) \rangle$ expressed by the following regression
\begin{equation}\label{Eq:Significance:Test}
  Y_t = c + aX_{t-\langle x(t)\rangle} + \varepsilon_t ~.
\end{equation}
The test consists in verifying that the coefficient $a$ should be different from 0 with high statistical significance.

\section{Interaction patterns between CNY-CNH exchange rates}
\label{S1:Interaction:patterns}

Let us now examine the time-dependent interaction relationship between the CNY and CNH exchange rates using the TOP method. We first analyze the daily rates.

\subsection{TOP analysis on daily exchange rates}

\begin{figure}[!htb]
  \centering
  \includegraphics[width=7.6cm, height=4.4cm]{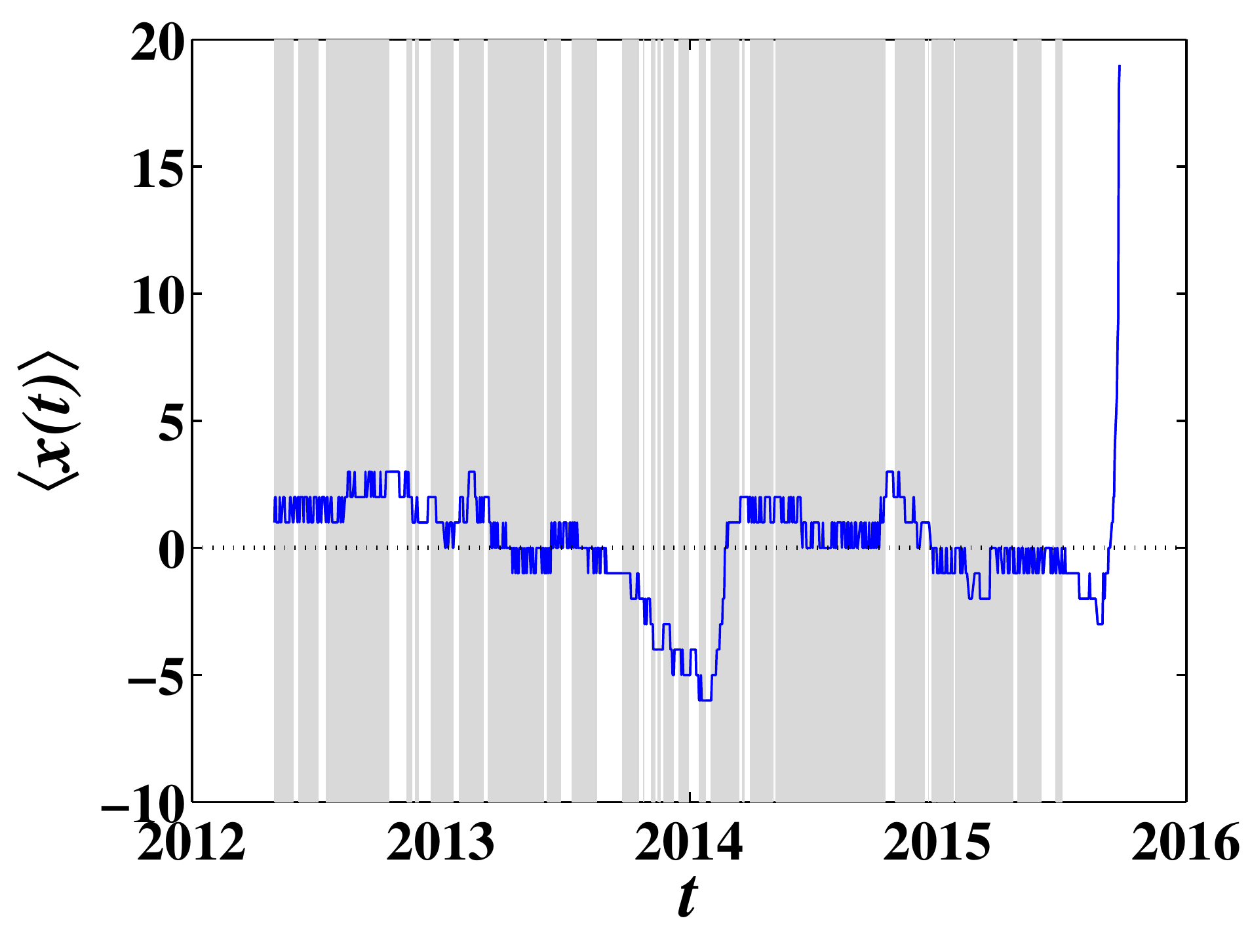}
  \includegraphics[width=7.6cm, height=4.4cm]{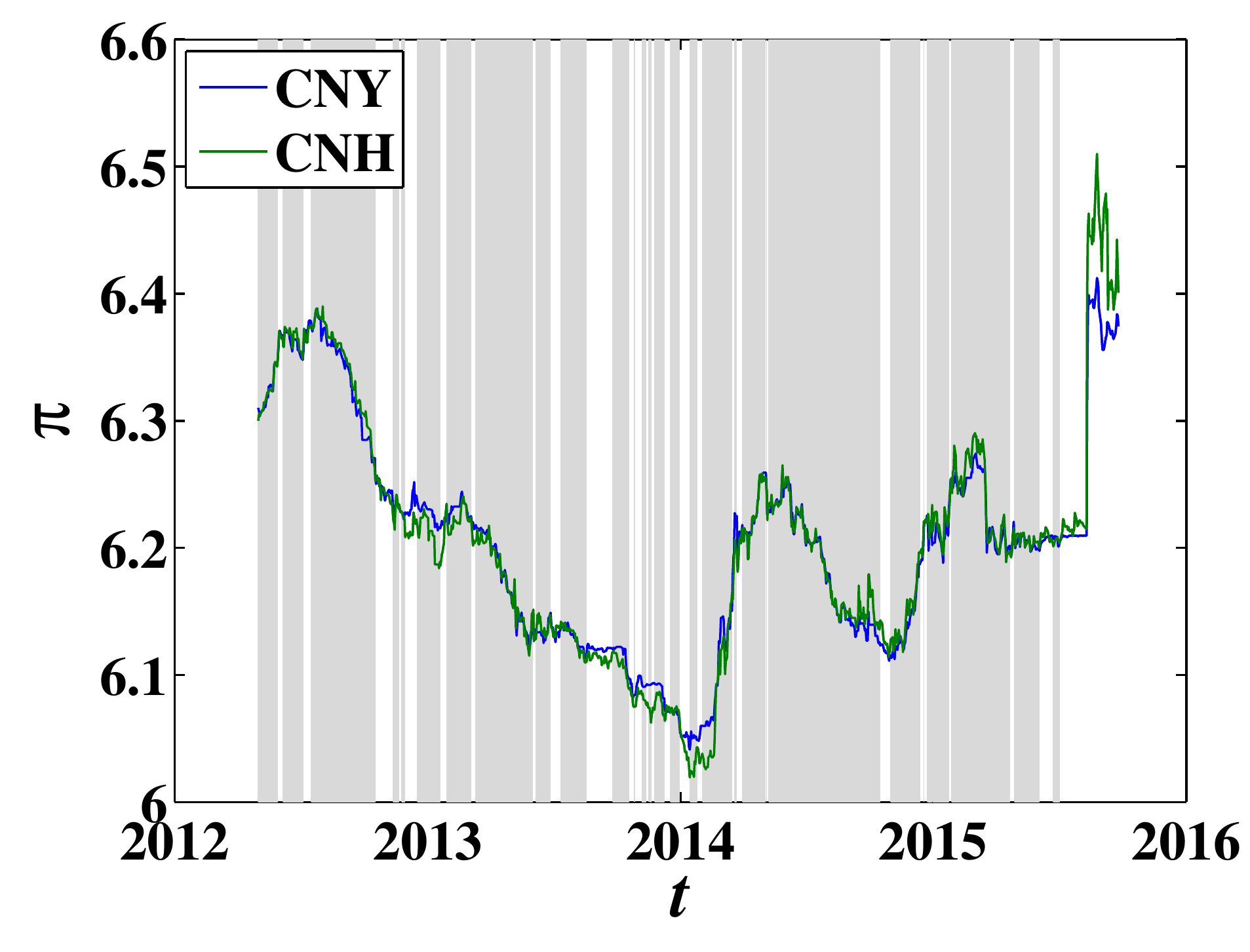}\\
  \includegraphics[width=0.33\linewidth]{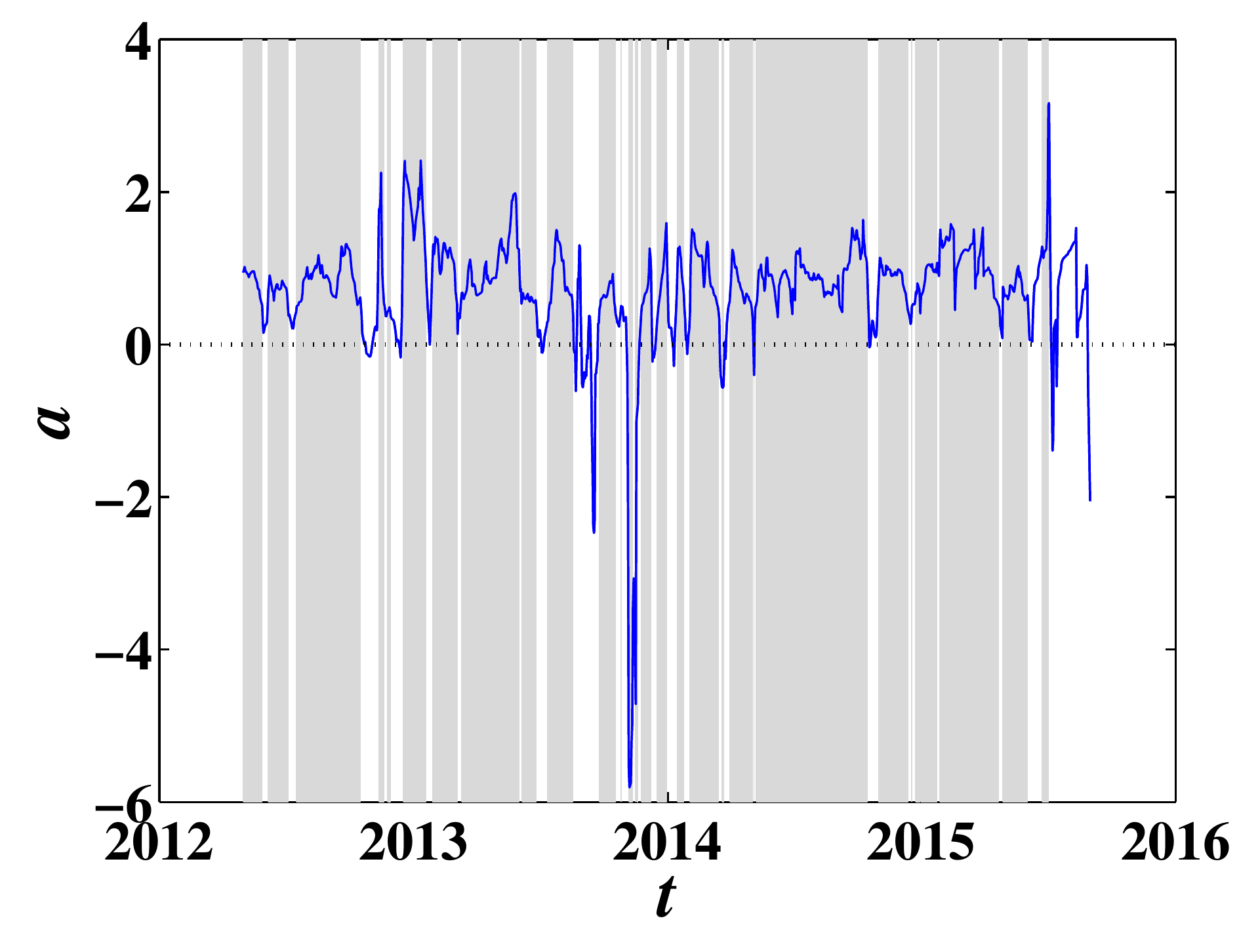}
  \includegraphics[width=0.33\linewidth]{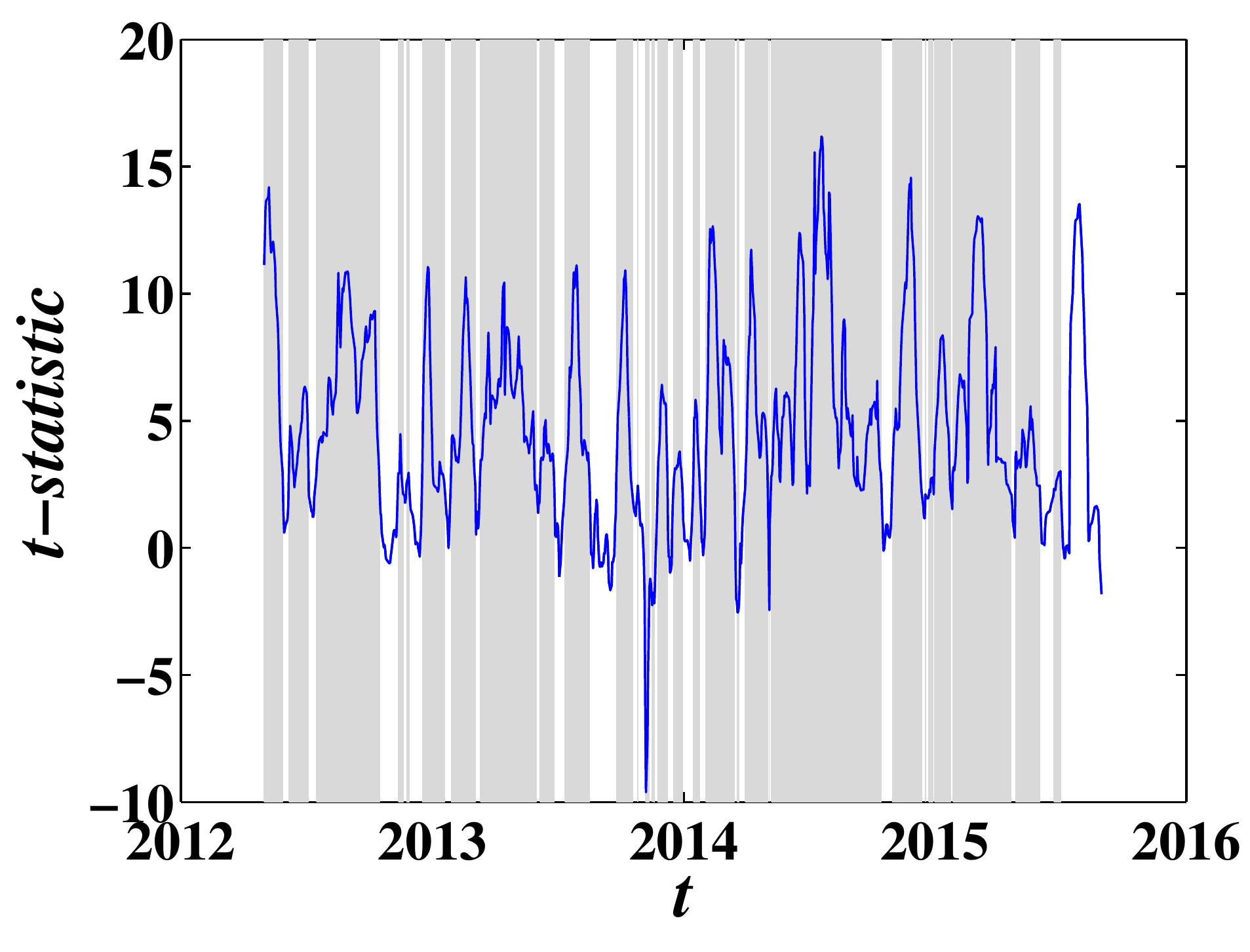}
  \includegraphics[width=0.33\linewidth]{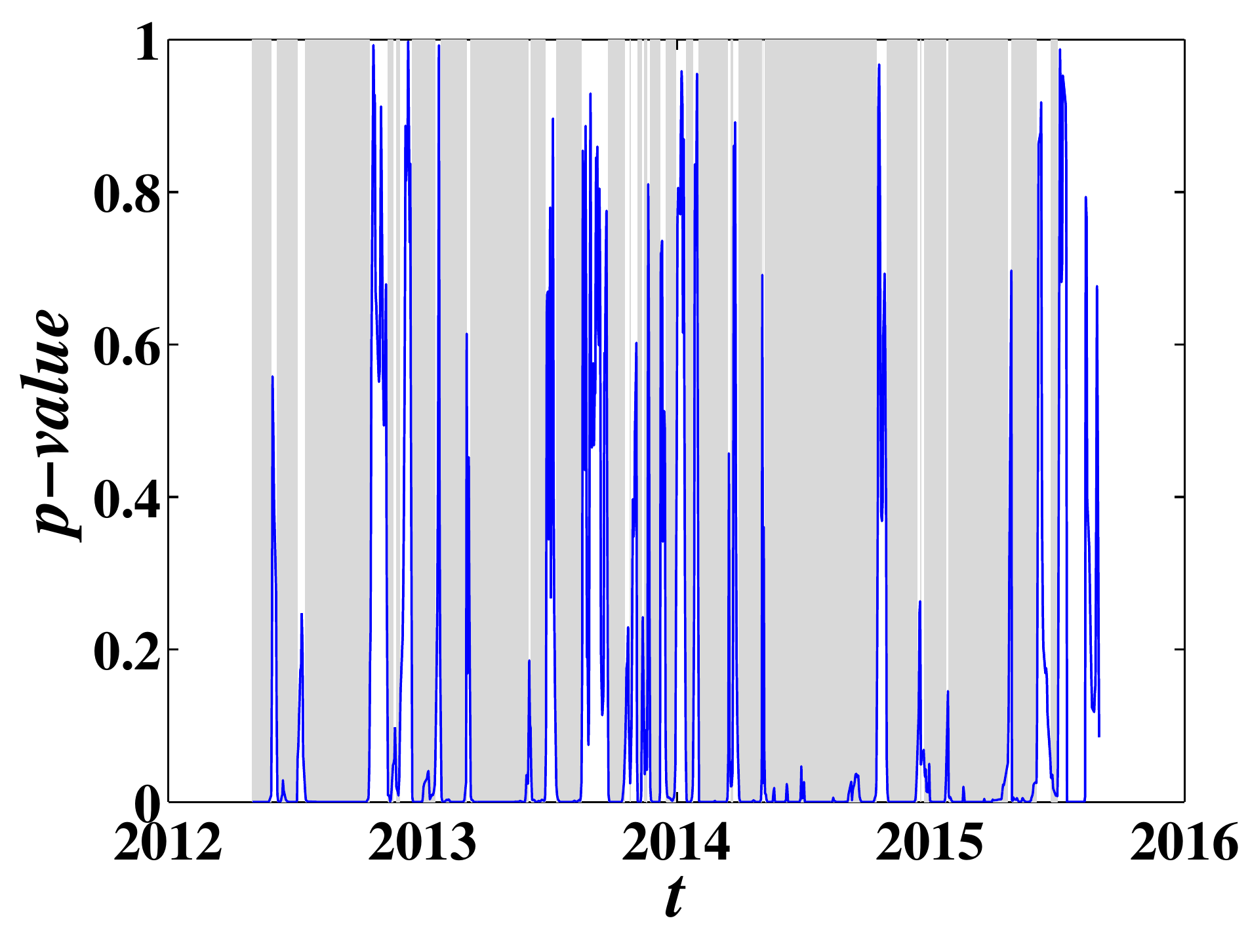}
  \vskip  -0.51\textwidth    \hskip   -0.92\textwidth (a)
  \vskip  -0.025\textwidth    \hskip   +0.04\textwidth (b)
  \vskip  +0.23\textwidth   \hskip   -0.96\textwidth (c)
  \vskip  -0.026\textwidth   \hskip   -0.3\textwidth (d)
  \vskip  -0.026\textwidth   \hskip   +0.36\textwidth (e)
  \vskip  +0.22\textwidth
  \caption{TOP analysis of the daily CNY and CNH exchange rates. The top left panel reports the average optimal thermal path $\langle x(t) \rangle$ at temperature $T=2$. The top right panel shows the corresponding CNY/USD (CNH/USD) exchange rates series $\pi$. The three bottom panels show respectively the time-dependence of coefficient $a$, its $t$-statistic and $p$-value of the consistency test (Eq.~\ref{Eq:Significance:Test}). The domains in grey indicate the times when the consistency test is significant at the 5\% level.}
  \label{Fig:TOP:Daily:2:20}
\end{figure}

Fig.~\ref{Fig:TOP:Daily:2:20} presents the average optimal thermal path between daily onshore and offshore exchange rates. We take 39 starting points and 39 ending points for TOP analyses, which are ($t_1=i,t_2=1$), ($t_1=1,t_2=i$), ($t_1=N-i,t_2=N$), ($t_1=N,t_2=N-i$) and ($t_1=N,t_2=N$), where $i = 1,2,\ldots,20$ and $N$ is the length of time series. The path with the minimum average energy defined by Eq.~(\ref{Eq:Energy}) over these 39 $\times$ 39 paths is the thermal optimal path we are looking for. A positive (resp. negative) $\langle x(t) \rangle$ indicates that the CNY rate leads (resp. lags) the CNH rate. We can observe that during most times $\langle x(t) \rangle$ lies within the interval $[-2,3]$, i.e. near the zero line, suggesting a weak lead-lag relationship between the CNY rate and the CNH rate. The price discovery process seems to alternate, with sometimes the CNY rate leading the CNH rate, while at other times
the CNH rate leads the CNY rate. However, around the beginning of 2014 and for about one year, $\langle x(t) \rangle$ became significantly lower than zero, indicating that the offshore exchange rate led the onshore exchange rate over this period.

In order to better understand the evolution of $\langle x(t) \rangle$, we also plot the CNY/USD (CNH/USD) exchange rates series $\pi$. In most cases, CNY and CNH exchange rates track each other quite well, which is in agreement with the weakly alternate lag structure we have obtained. However, there are episodes in which CNY and CNH display a large difference. For instance, in January 2014, CNH appreciates at an accelerating rate and gradually separates with CNY. Then CNH depreciates rapidly and narrows the gap with CNY in February. This big turning point reflects the transformation of offshore investors' expectations. After a whole year's appreciation of the Renminbi in 2013, investors believed that this trend will continue at first, but then, with the recession of China's economy and the influence of other macro factors, offshore investors increased their expectations of a devaluation and closed their positions on
the Renminbi. As a result, the offshore exchange rate exerted an influence on the onshore exchange rate.

Another large disparity between CNY/USD and CNH/USD that should be mentioned happens at the end of our sample. Our TOP analysis shows that $\langle x(t) \rangle$ is shooting up very fast to large positive values, indicating that the CNY exchange rate precedes the CNH exchange rate at this time. This phenomenon can be explained by the fact that, on 11 August 2015, the People's Bank of China actively depreciated the central parity of Renminbi to 6.2298. Hence, this price change was introduced by the onshore market
and then cascaded to the offshore market.

We also report in Fig.~\ref{Fig:TOP:Daily:2:20}(c-e) the time dependence of coefficient $a$, its $t$-statistic and the $p$-value of the consistency test defined by equation Eq.~(\ref{Eq:Significance:Test}). The grey shades in all plots indicate that the
consistency test is significant at the 5\% level ($p\leq0.05$). Each test is a regression with time horizon of 20 days. We can observe that, most of the time, the consistency test is significant, with $a$ deviating from 0 and the absolute value of the $t$-statistic being greater than 2. We also notice that, except for the two episodes of large price disparities, the coefficient $a$ almost always keeps positive
with a typical value fluctuating around 1.

\begin{figure}[t!b]
  \centering
  \includegraphics[width=7.6cm, height=4.4cm]{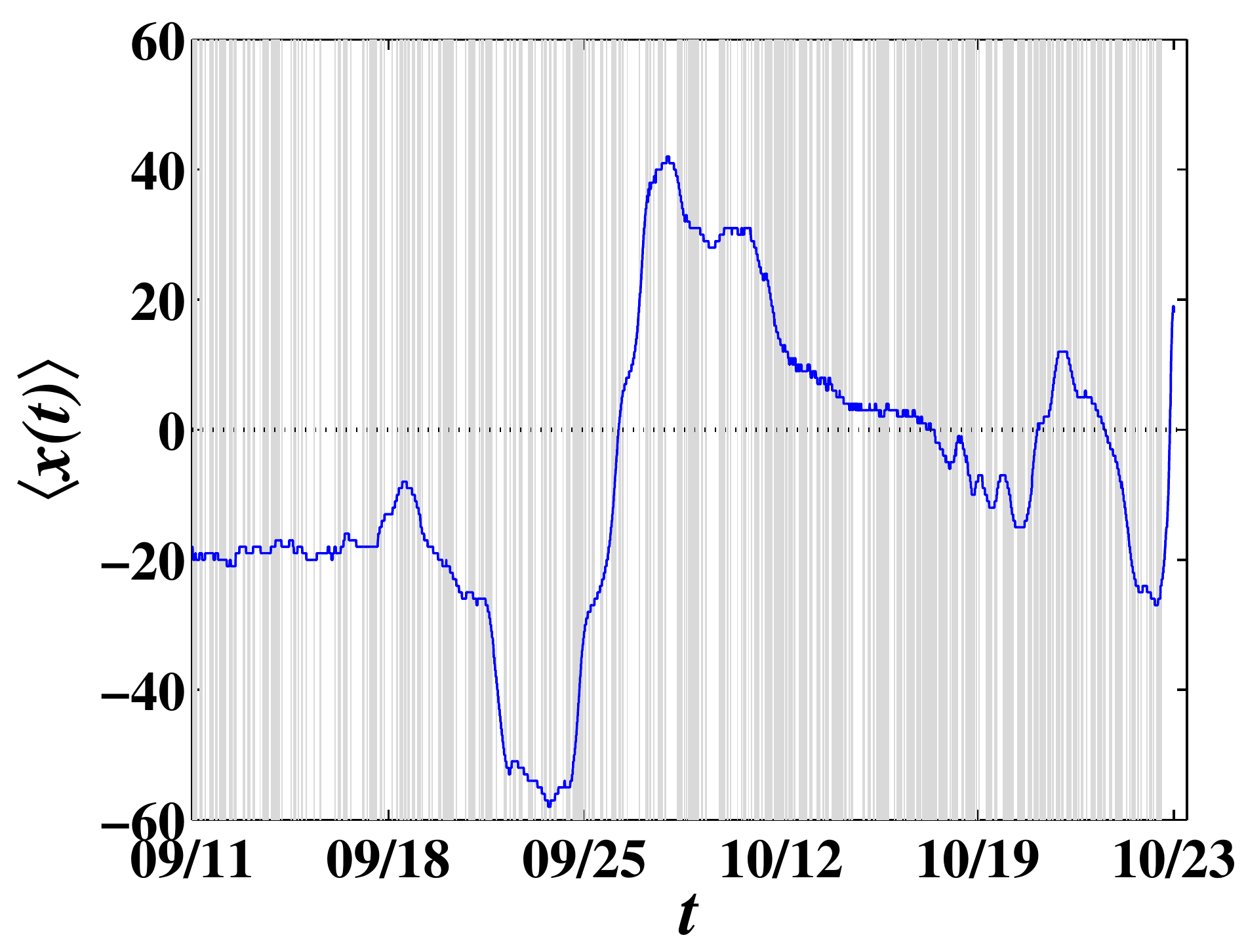}
  \includegraphics[width=7.6cm, height=4.4cm]{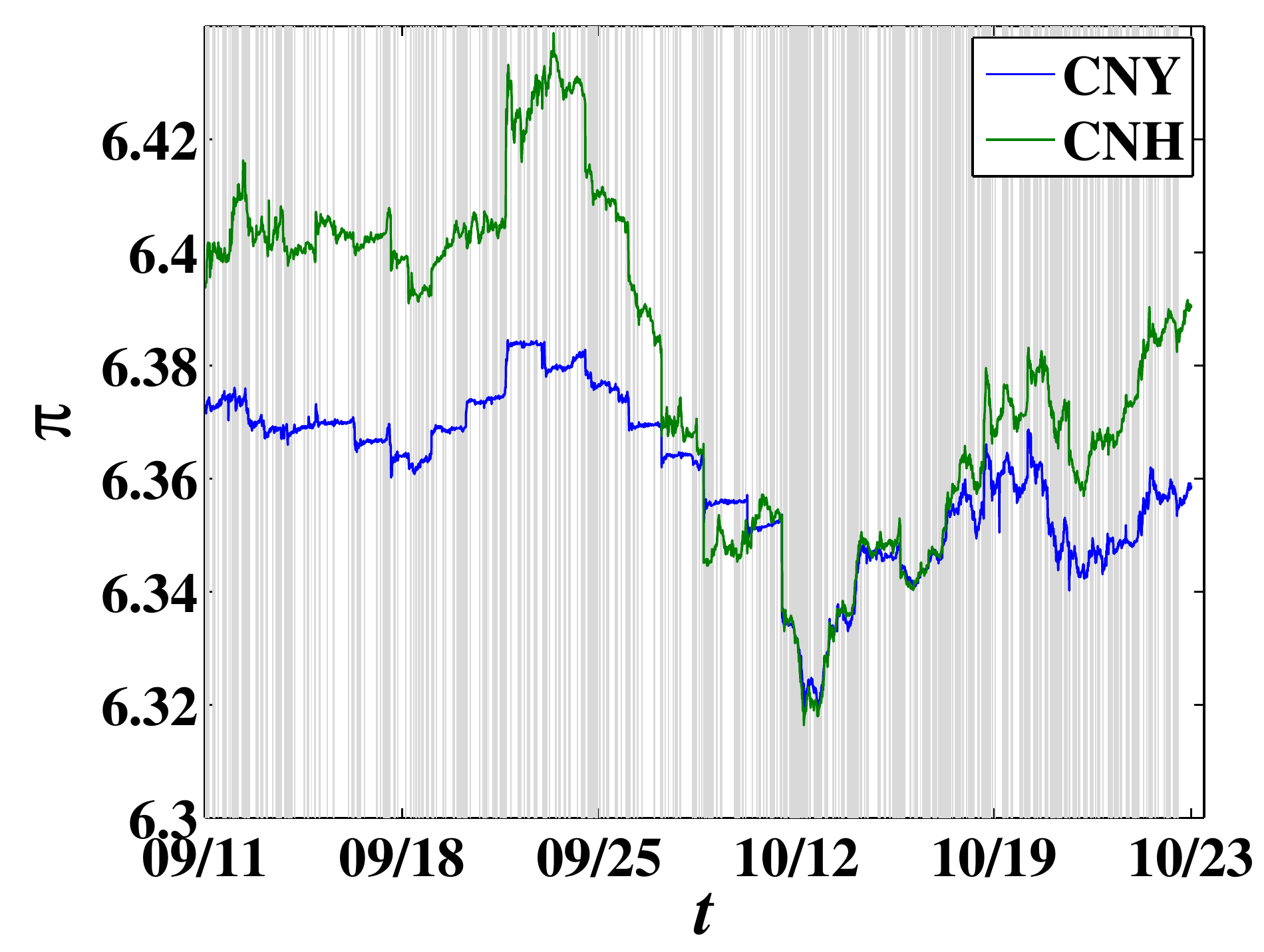}\\
  \includegraphics[width=0.33\linewidth]{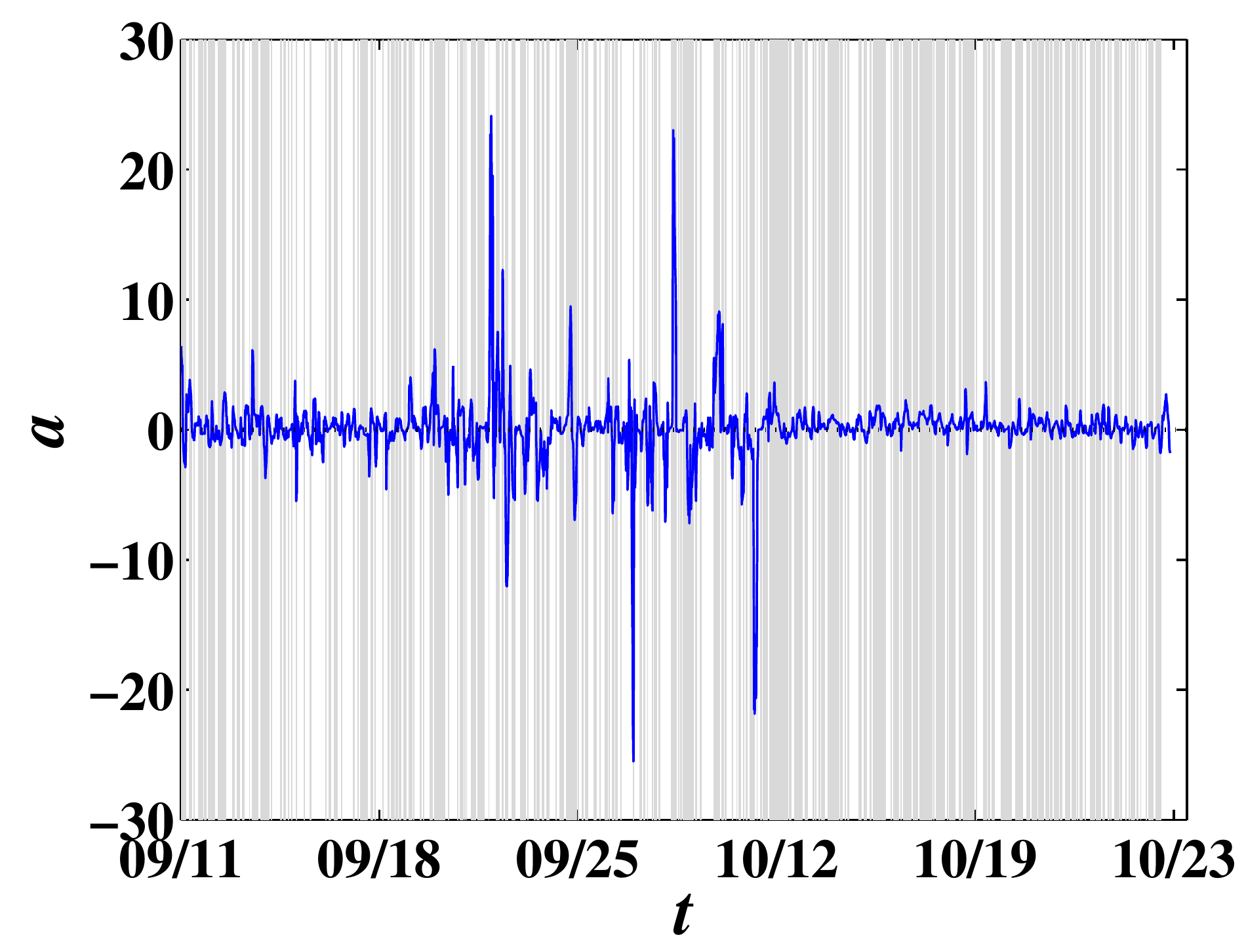}
  \includegraphics[width=0.33\linewidth]{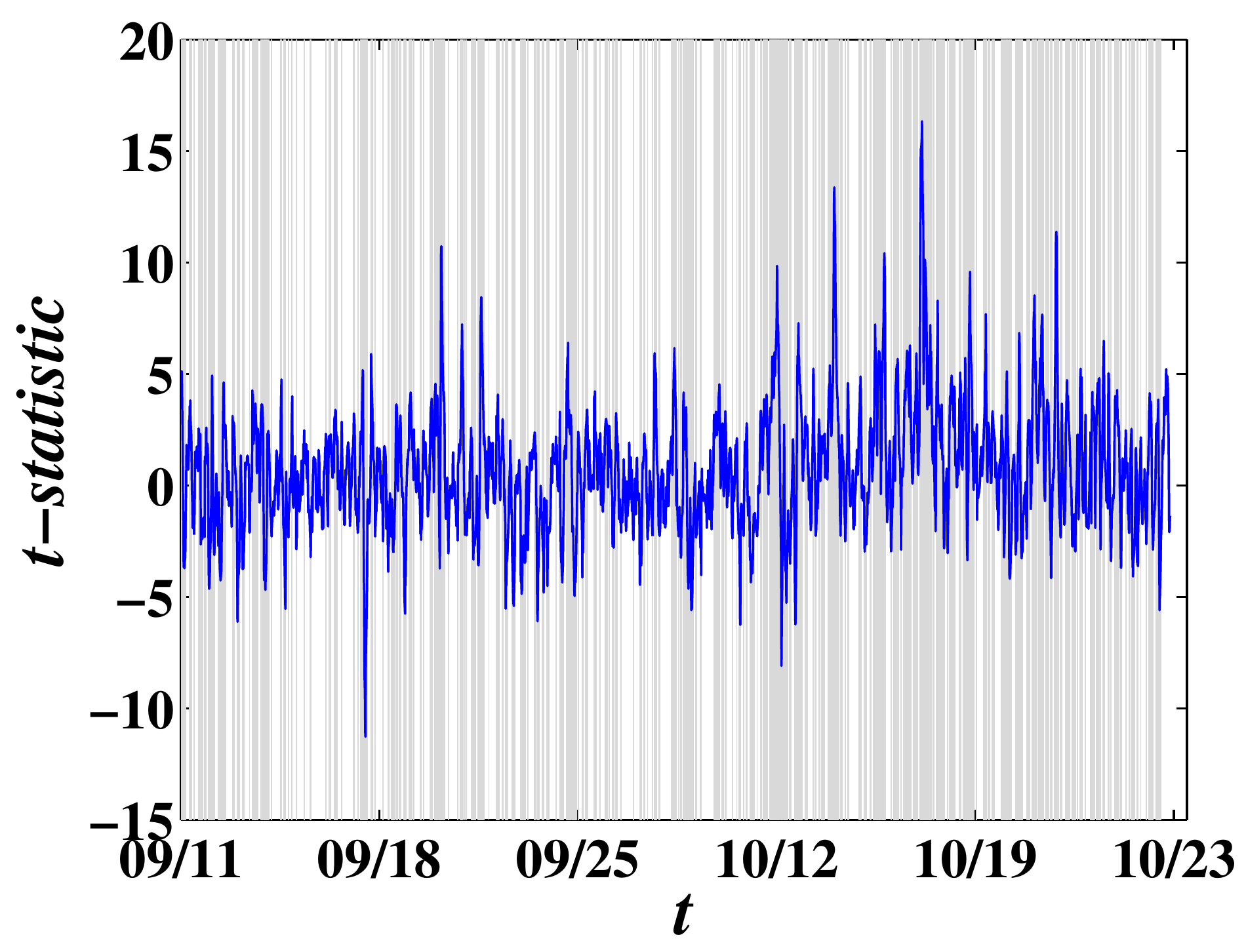}
  \includegraphics[width=0.33\linewidth]{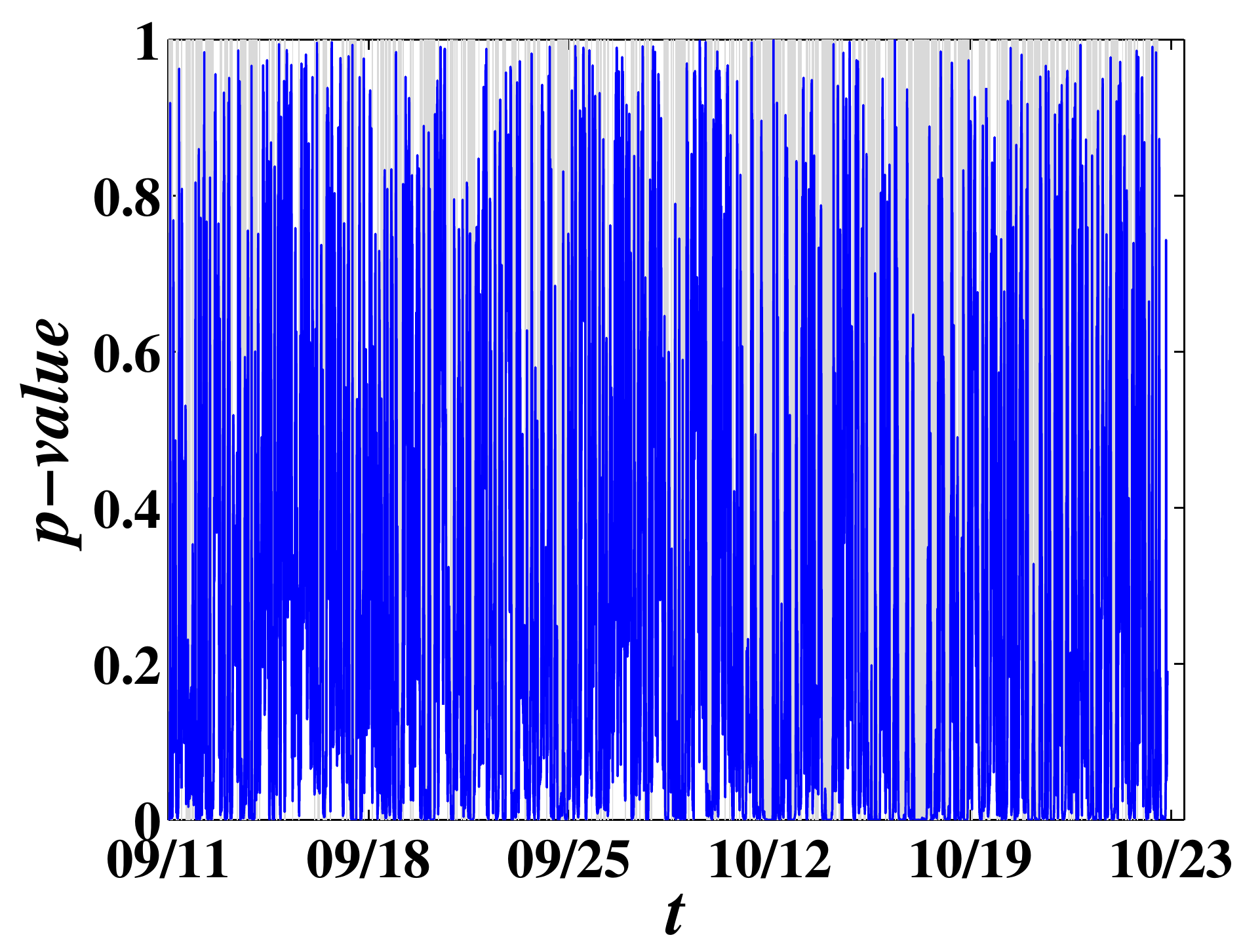}
  \includegraphics[width=0.33\linewidth]{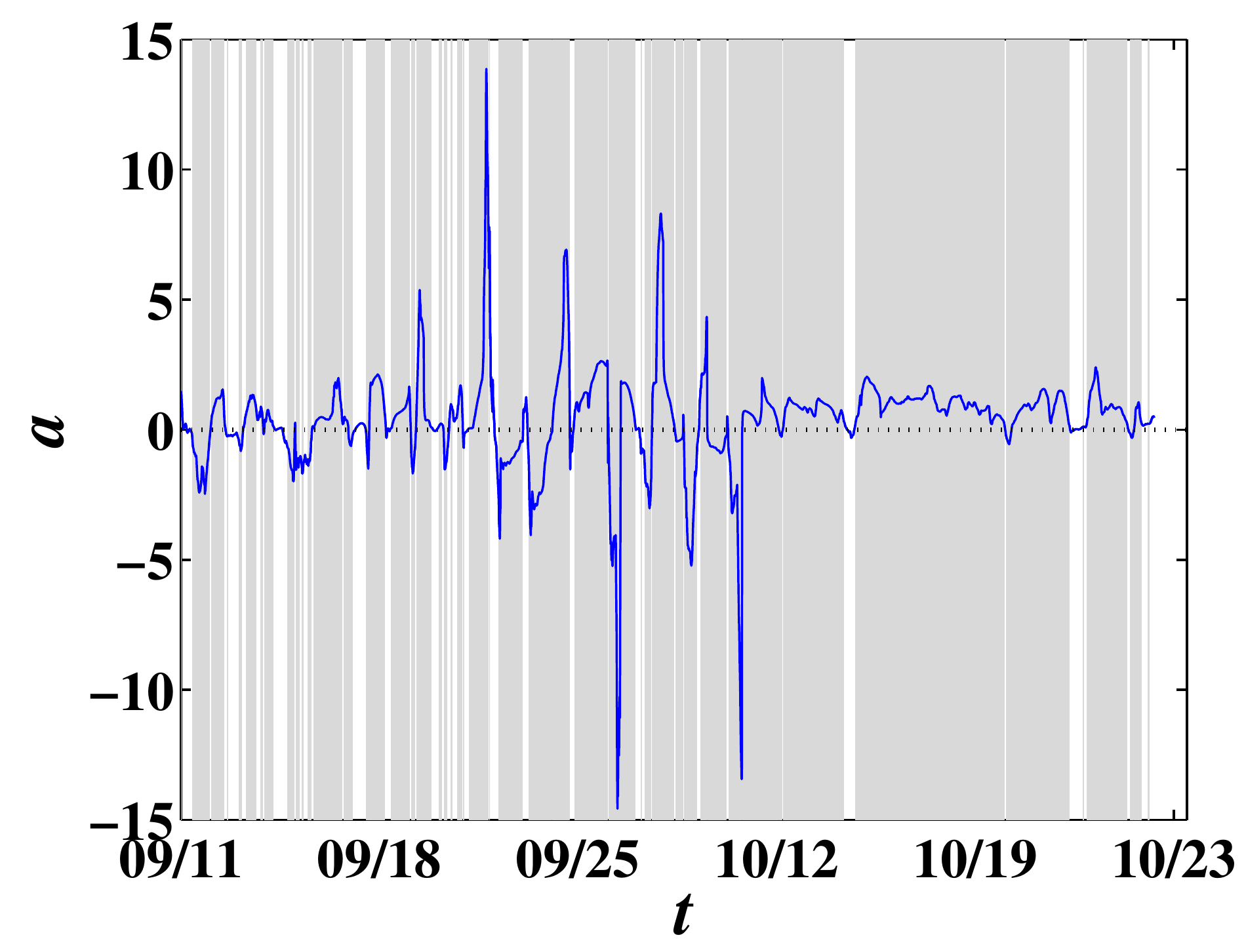}
  \includegraphics[width=0.33\linewidth]{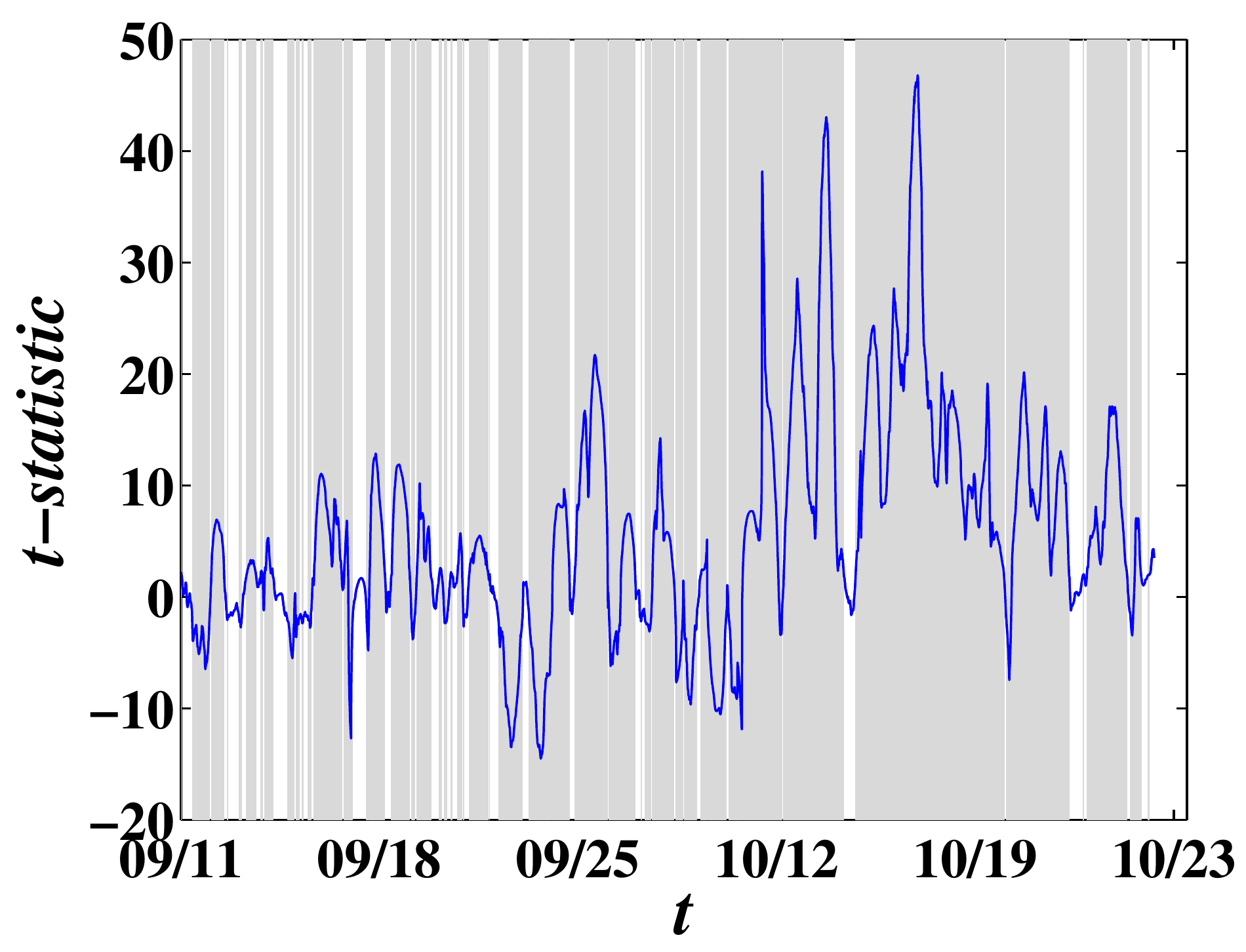}
  \includegraphics[width=0.33\linewidth]{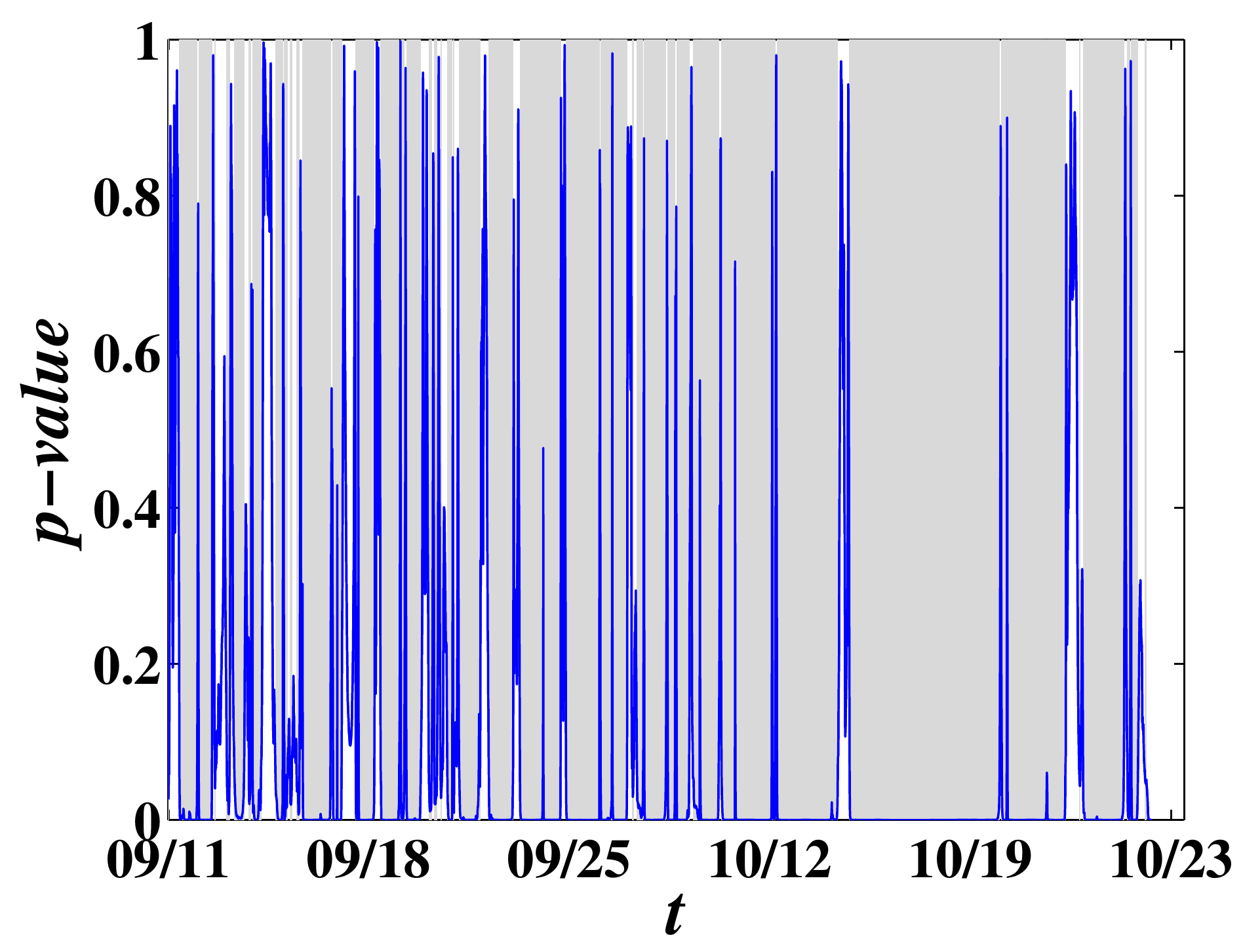}
  \vskip  -0.77\textwidth    \hskip   -0.92\textwidth (a)
  \vskip  -0.025\textwidth    \hskip   +0.04\textwidth (b)
  \vskip  +0.23\textwidth   \hskip   -0.96\textwidth (c)
  \vskip  -0.026\textwidth   \hskip   -0.3\textwidth (d)
  \vskip  -0.026\textwidth   \hskip   +0.36\textwidth (e)
  \vskip  +0.23\textwidth   \hskip   -0.96\textwidth (f)
  \vskip  -0.026\textwidth   \hskip   -0.3\textwidth (g)
  \vskip  -0.026\textwidth   \hskip   +0.36\textwidth (h)
  \vskip  +0.22\textwidth
  \caption{TOP analysis of the minute-scale CNY and CNH exchange rates from 11 September 2015 to 23 October 2015. The grey shades represent the periods when the consistency test is significant at the 5\% level. (a) Average optimal thermal path $\langle x(t) \rangle$ at temperature $T=2$. (b) CNY/USD (CNH/USD) exchange rates series $\pi$ at the one minute time scale as a function of time. (c) Time dependence of coefficient $a$ of the consistency test in Eq.~(\ref{Eq:Significance:Test}) estimated in running windows of 20 minutes duration. (d) Corresponding $t$-statistic estimated in time windows of 20 minutes duration. (e) Corresponding $p$-values. (f) Time dependence of coefficient $a$ of the consistency test defined by equation Eq.~(\ref{Eq:Significance:Test}) in 100-min windows. (g) Corresponding $t$-statistic. (h) Corresponding $p$-values.}
  \label{Fig:TOP:Minutely:2:20:100}
\end{figure}

\subsection{TOP analysis for minute-scale exchange rates}

Fig.~\ref{Fig:TOP:Minutely:2:20:100}(a) presents the average optimal thermal path between minute-scale onshore and offshore exchange rates. Totally different from the interaction pattern between daily exchange rates, the lead-lag path $\langle x(t) \rangle$ for minute-scale CNY-CNH exchange rates exhibits a richer dynamics, departing significantly from zero most of the time.
This can be associated with the fact that the price series of CNY and CNH no longer track each other well and large differences
between them appear at many times. The lead-lag path $\langle x(t) \rangle$ is negative at most times, meaning that, at short time scales, the offshore exchange rate often plays a leading role and influences the onshore exchange rate.
One can also observe the occurrence of a few periods with positive $\langle x(t) \rangle$. For instance, from 28 September to 12 October 2015, the onshore exchange rate precedes the offshore exchange rate in their lead-lag relationship. It is worth noting that this episode corresponds to a downward trending market. From 12 October to 19 October when the two exchange rates are very close
to each other, the average optimal thermal path $\langle x(t) \rangle$ is correspondingly found to be close to $0$.

Due to the larger fluctuations and disparity between minute-scale CNY-CNH exchange rates, the results of the consistency tests
shown in Fig.~\ref{Fig:TOP:Minutely:2:20:100}(c-e) give more insignificant $a$ values when compared to the daily rates. However,
for most of the time windows, the coefficient $a$ is qualified as significantly non-zero with a small $p$-value, as in panels (e) and (h).
For window lengths of 100 minutes, the test statistics become much cleaner, as shown in Fig.~\ref{Fig:TOP:Minutely:2:20:100}(f-h).

\begin{figure}[!tb]
  \centering
  \includegraphics[width=0.33\linewidth]{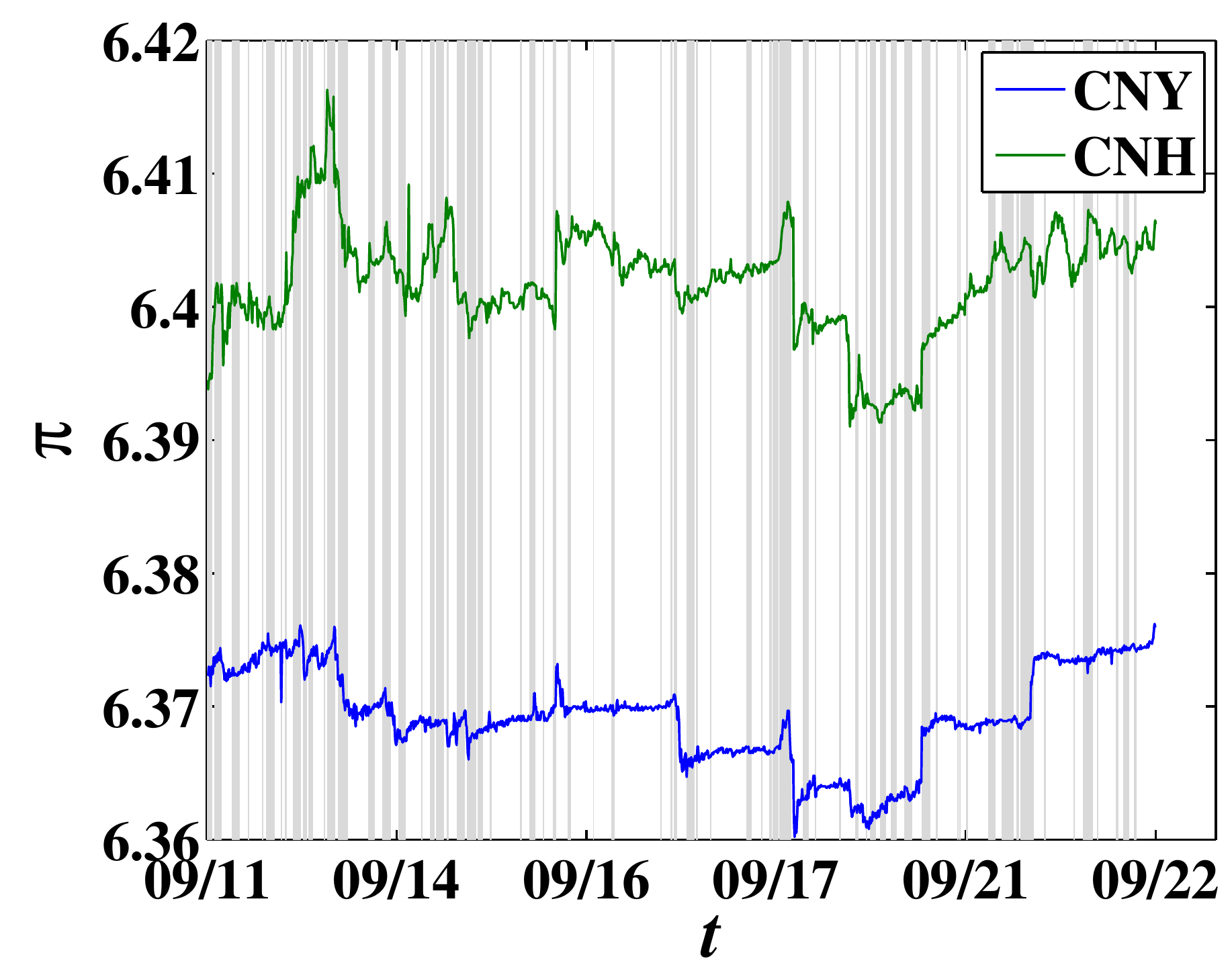}
  \includegraphics[width=0.33\linewidth]{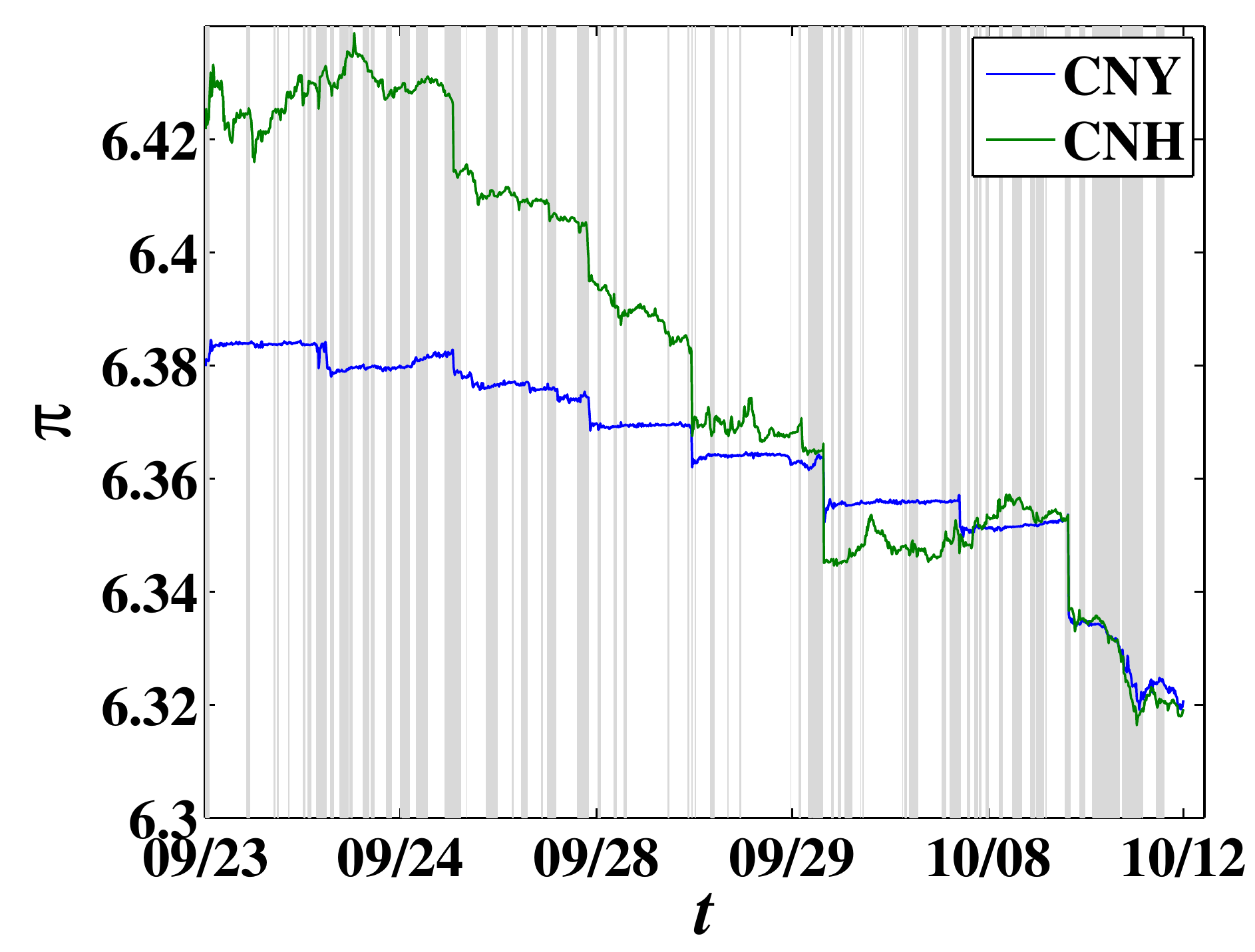}
  \includegraphics[width=0.33\linewidth]{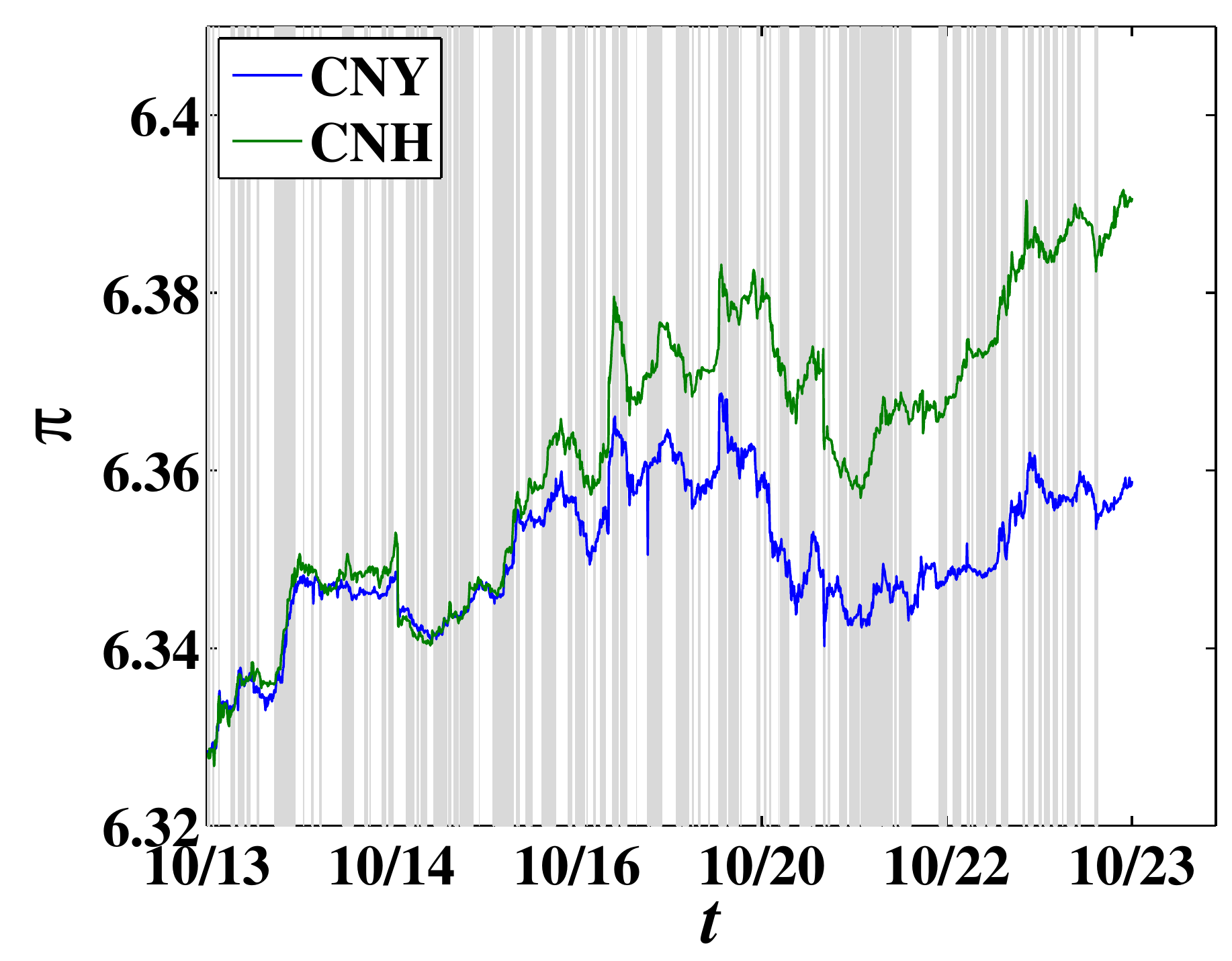}\\
  \includegraphics[width=0.33\linewidth]{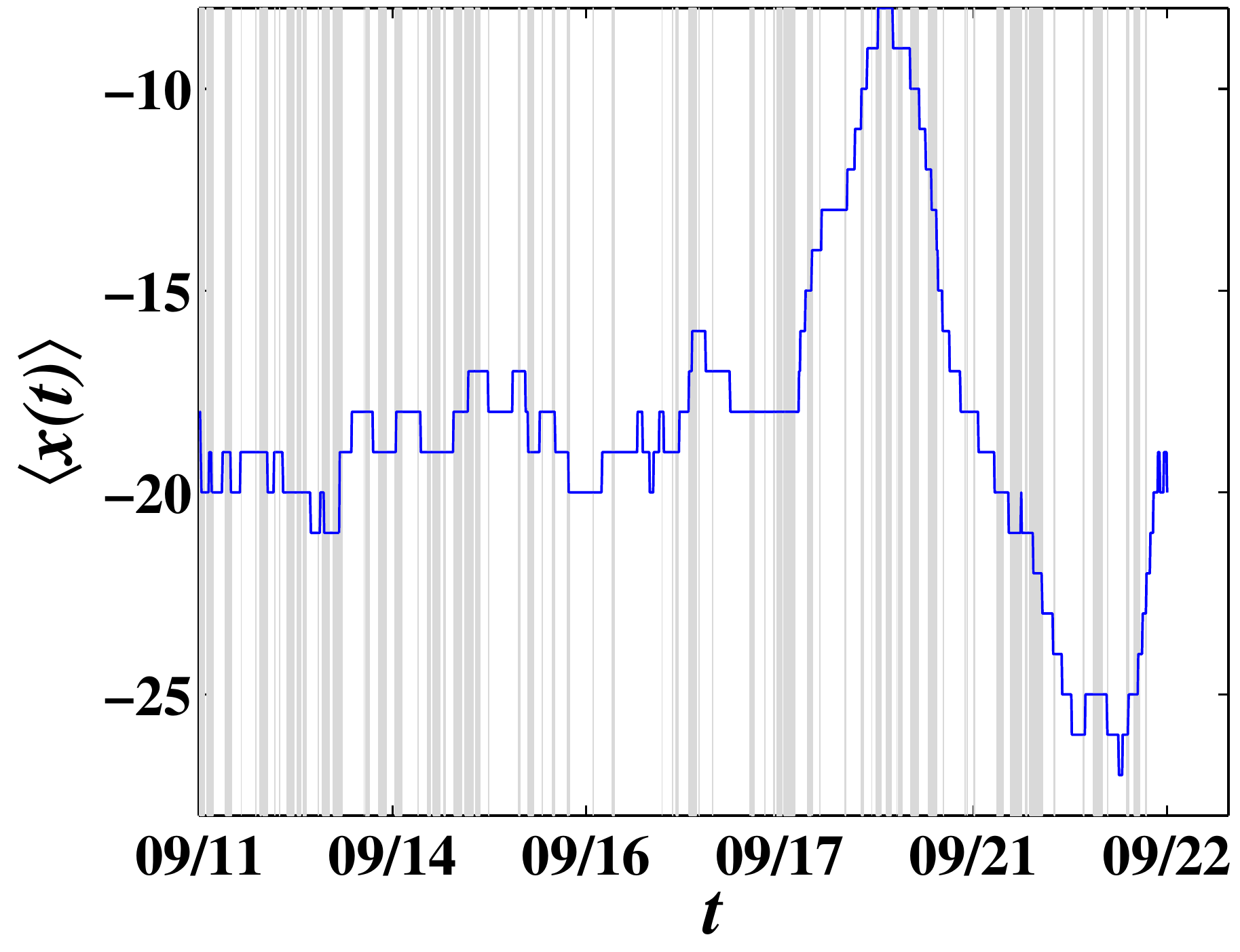}
  \includegraphics[width=0.33\linewidth]{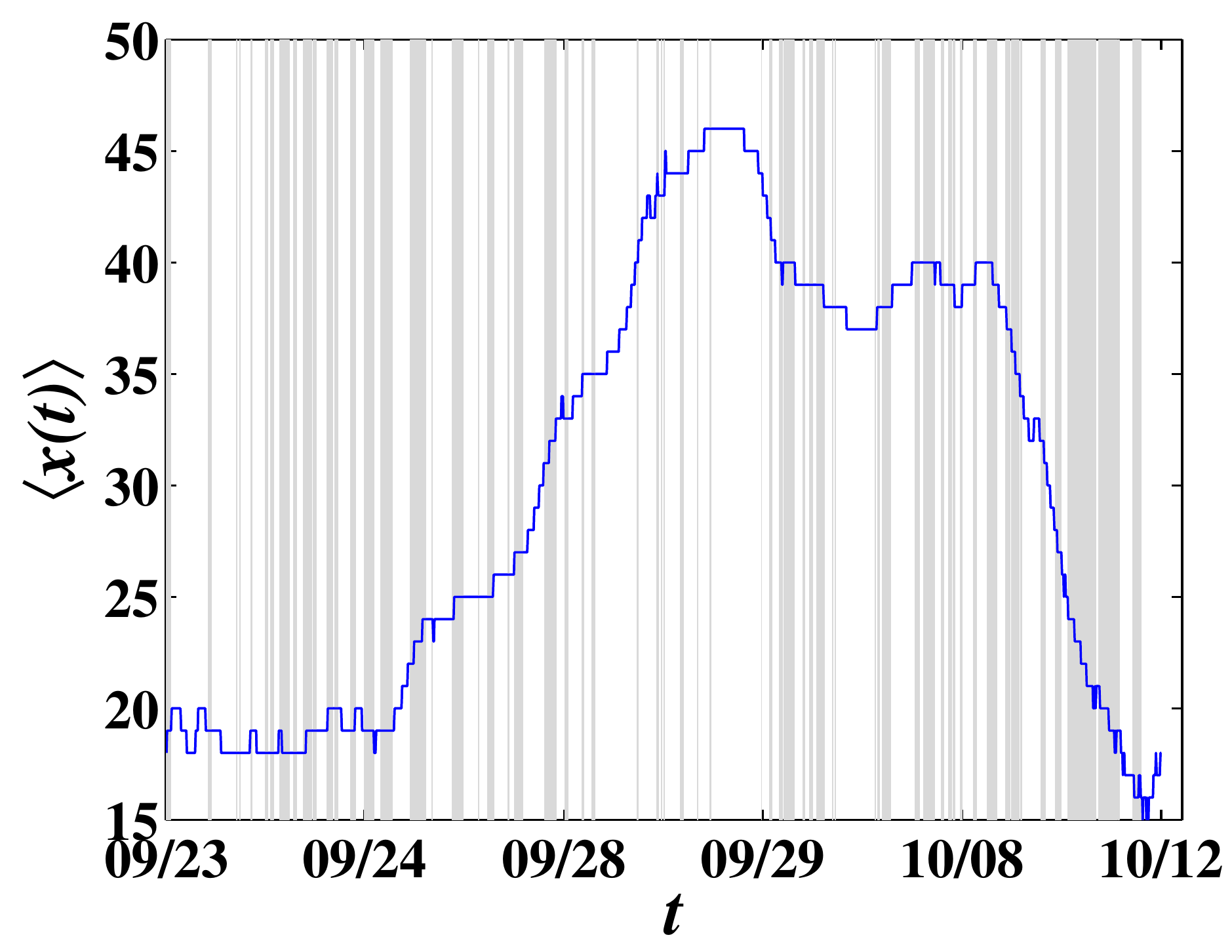}
  \includegraphics[width=0.33\linewidth]{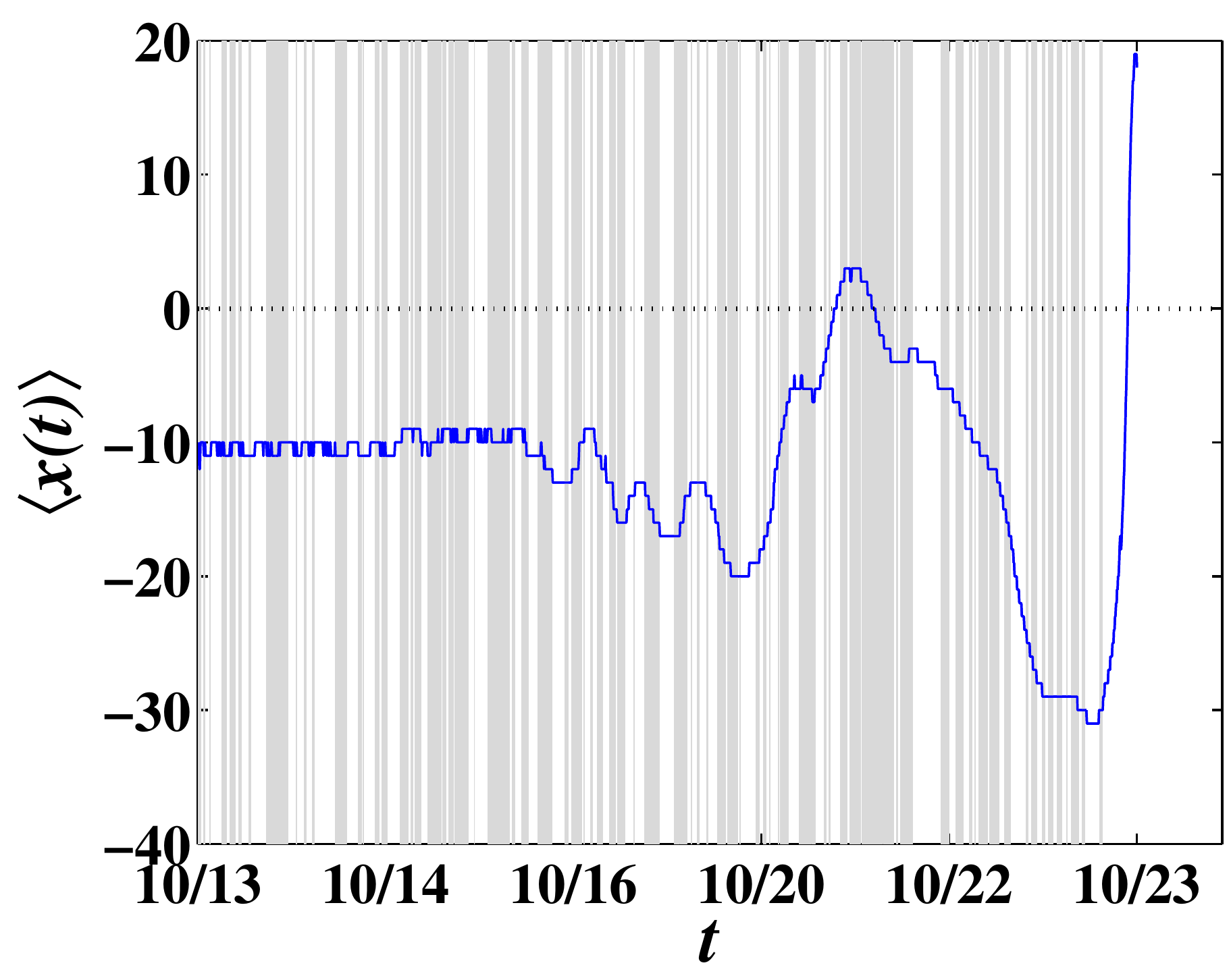}\\
  \includegraphics[width=0.33\linewidth]{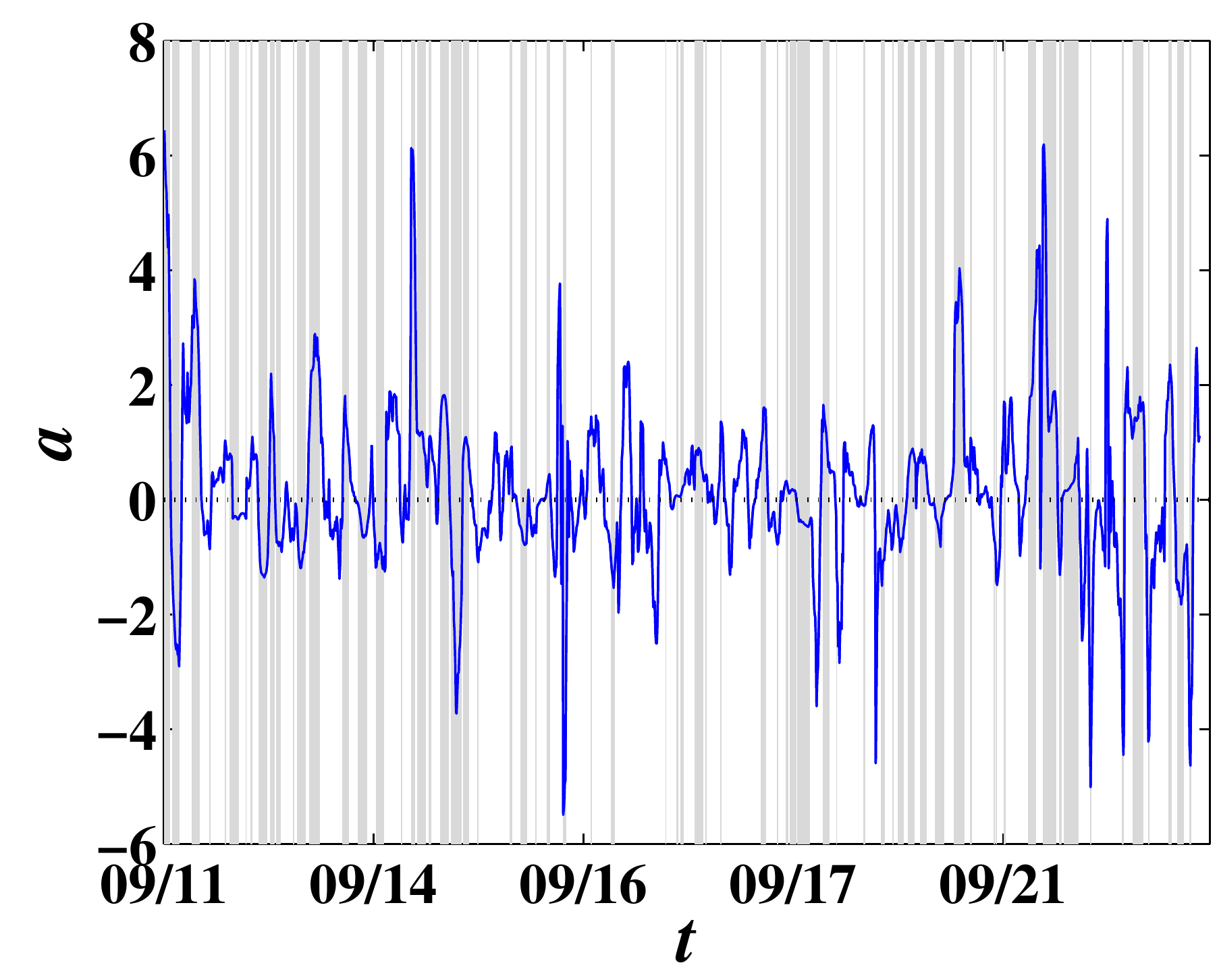}
  \includegraphics[width=0.33\linewidth]{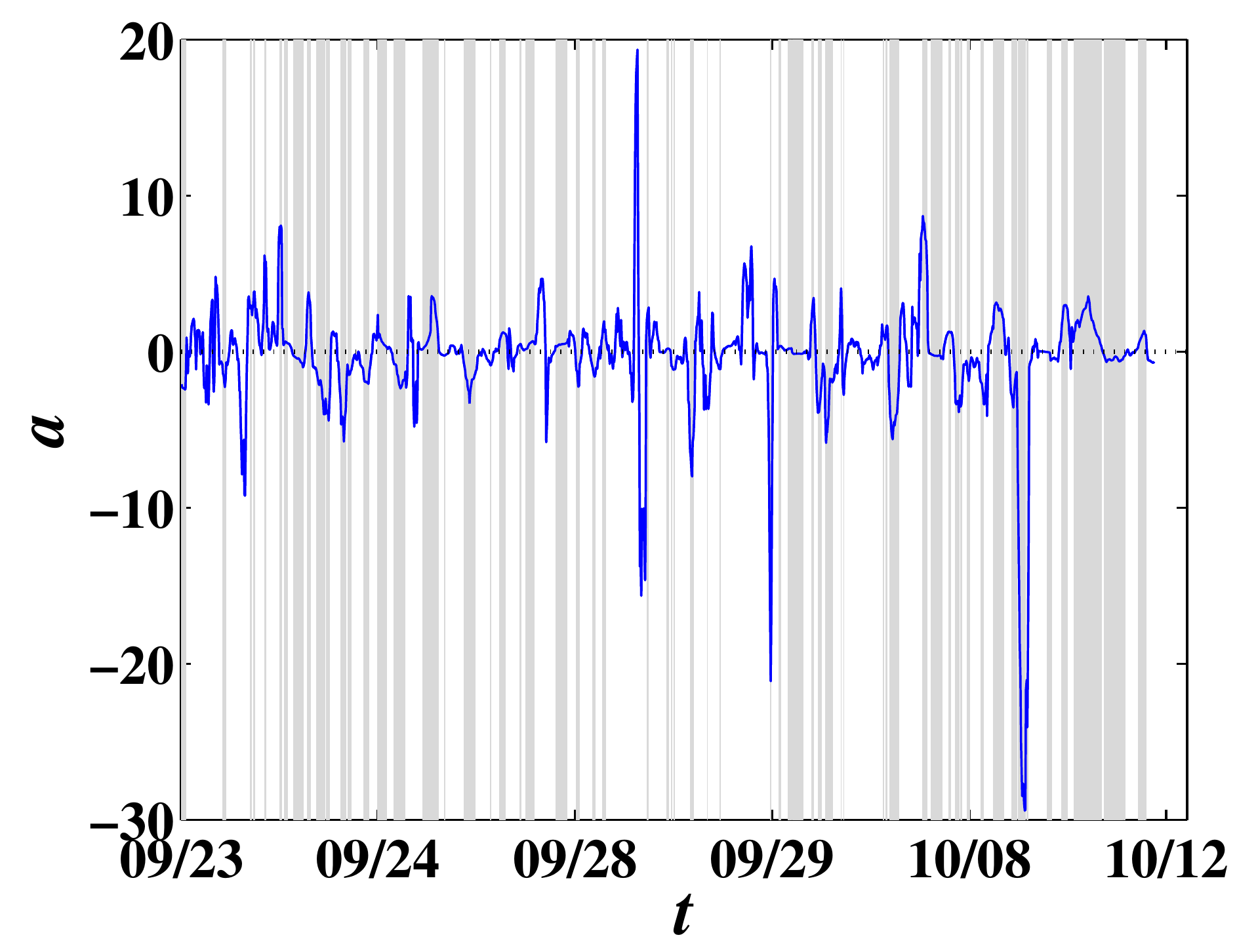}
  \includegraphics[width=0.33\linewidth]{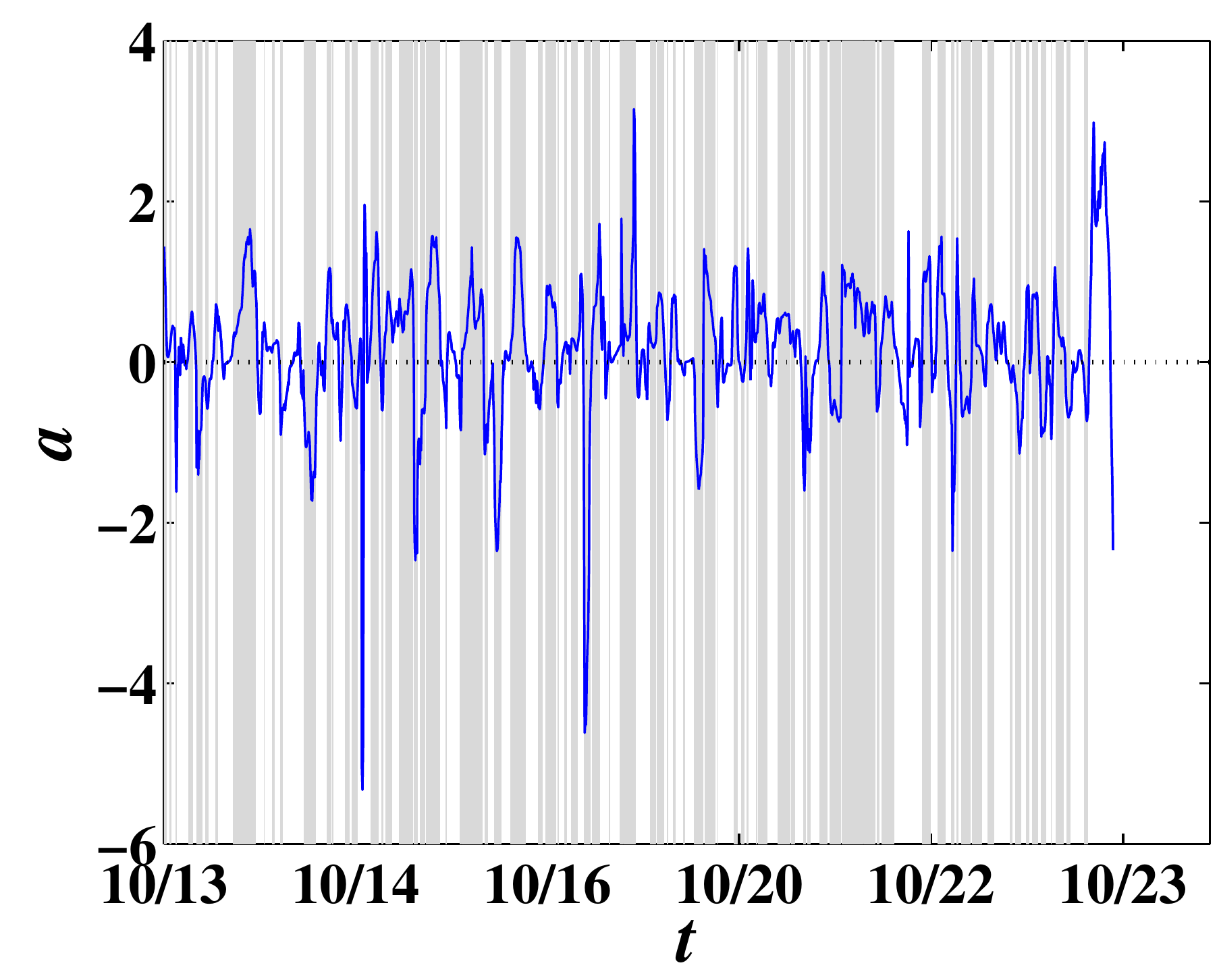}\\
  \vskip  -0.782\textwidth    \hskip   -0.97\textwidth (a)
  \vskip  -0.026\textwidth    \hskip  -0.3\textwidth (b)
  \vskip  -0.026\textwidth    \hskip  +0.36\textwidth (c)
  \vskip  +0.23\textwidth    \hskip   -0.97\textwidth (d)
  \vskip  -0.026\textwidth   \hskip   -0.3\textwidth (e)
  \vskip  -0.026\textwidth   \hskip   +0.36\textwidth (f)
  \vskip  +0.23\textwidth    \hskip   -0.97\textwidth (g)
  \vskip  -0.026\textwidth   \hskip   -0.3\textwidth (h)
  \vskip  -0.026\textwidth   \hskip   +0.36\textwidth (i)
  \vskip  +0.24\textwidth
  \caption{TOP analysis of the minute-scale CNY and CNH exchange rates in three sub-periods. Panels (a-c) show
  respectively the two exchange rates in the three sub-periods, from 11 September 2015 to 22 September 2015 (panel (a)
  showing stable exchange rates),  from 23 September 2015 to 12 October 2015 (panel (b) showing a downward trending market)
  and from 13 October 2015 to 23 October 2015 (panel (c) showing an upward trending market).
 The second row of panels (d-f) show the average optimal thermal path $\langle x(t) \rangle$ at temperature $T=2$. The third row
 of panels (g-i) report the time-dependence of coefficient $a$. Grey shade domains indicate the time periods when the
consistency test is significant at the 5\% level. }
  \label{Fig:TOP:Sub:2:20}
\end{figure}

\subsection{TOP analysis with three sub-samples}

As a robustness check, we divide the whole minute-scale sample into three sub-samples to study the time-dependence of
the lead-lag relationship between the CNY and CNH exchange rates by performing the TOP analysis for each sub-sample independently. The first sub-sample contains data from 11 September 2015 to 22 September 2015, corresponding to a stable market state. The second sub-sample contains data from 23 September 2015 to 12 October 2015, corresponding to a downward trending market. The third sub-sample contains data from 13 October 2015 to 23 October 2015, corresponding to an upward trending market. The results are presented in Fig.~\ref{Fig:TOP:Sub:2:20}.

As a result of the different sample lengths and different starting or ending points, the average optimal thermal paths
determined in the three sub-samples deviat from the global optimal path for the whole sample shown in Fig.~\ref{Fig:TOP:Minutely:2:20:100}. However, the main characteristics and trends of $\langle x(t) \rangle$ in each sub-sample remain extremely similar with those of the whole sample path. For instance, in the first sub-period corresponding to a stable market state, the path $\langle x(t) \rangle$ first lies around -20 until 17 September, then quickly decreases to -9 on 18 September, and then reverses to lower than -25 on 22 September. The negative values of $\langle x(t) \rangle$ during this period indicates that the offshore exchange rate leads the onshore exchange rate. For the second sub-period corresponding to a downward trending market (i.e. an appreciation of the Renminbi versus US\$), both sub-sample path in Fig.~\ref{Fig:TOP:Sub:2:20} and whole sample path in Fig.~\ref{Fig:TOP:Minutely:2:20:100}
exhibit a positive hump shape from 28 September to 12 October. The positive values of $\langle x(t) \rangle$ confirm that the onshore exchange rate leads the offshore exchange rate during periods of Renminbi appreciation.
Finally, in the third sub-sample with an upward trending market, the lead-lag path $\langle x(t) \rangle$ shown in Fig.~\ref{Fig:TOP:Sub:2:20} is also very similar to the right-end part of $\langle x(t) \rangle$ shown in Fig.~\ref{Fig:TOP:Minutely:2:20:100}. The prevailing negative values of $\langle x(t) \rangle$ in this third sub-period confirms that the offshore exchange rate leads the onshore exchange rate in upward market state (i.e. during times of Renminbi depreciation).

\section{Alternative specifications}
\label{S1:Alternative:specifications}

\subsection{Alternative parameter $T$}

\begin{figure}[!htb]
  \centering
  \includegraphics[width=0.32\linewidth]{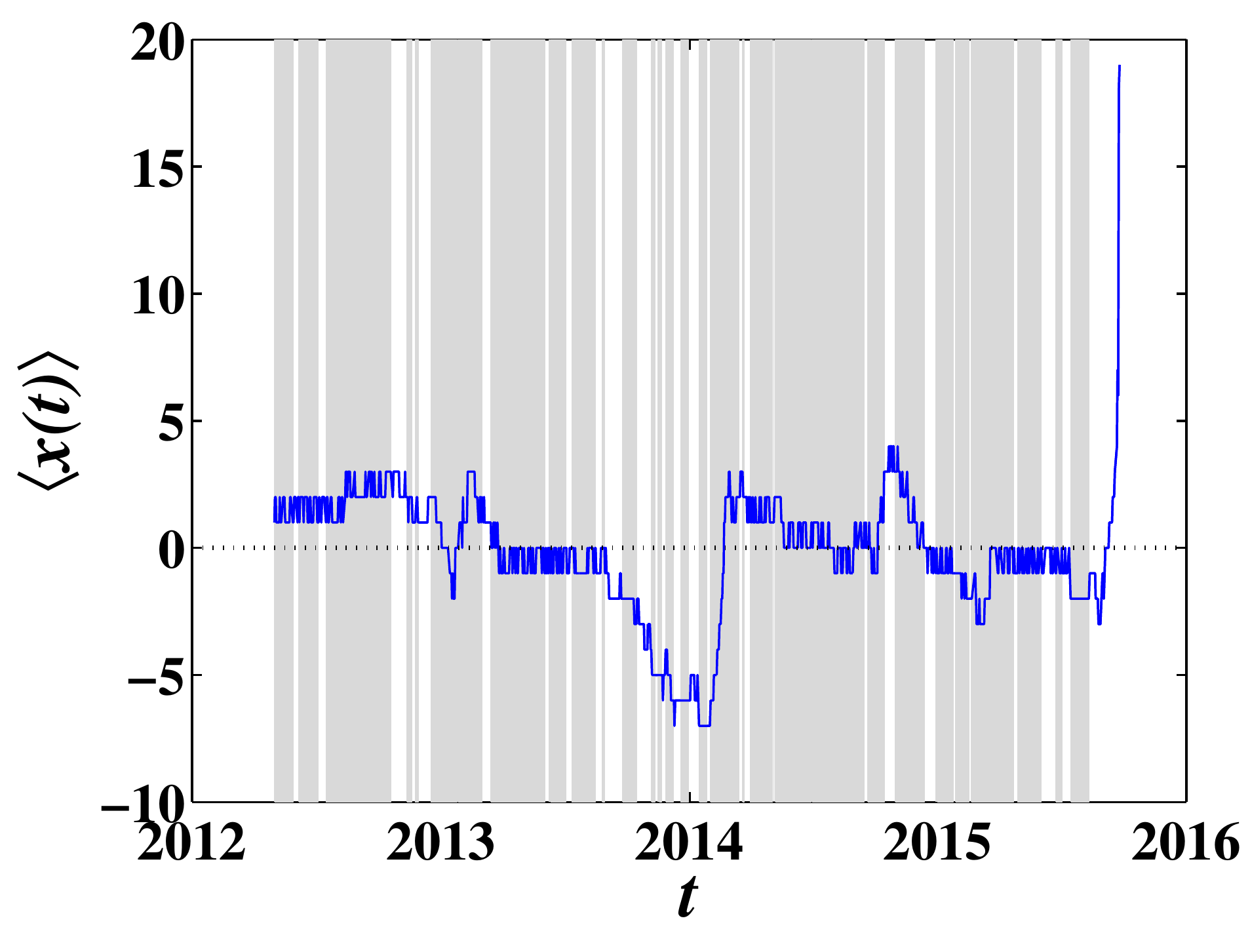}
  \includegraphics[width=0.32\linewidth]{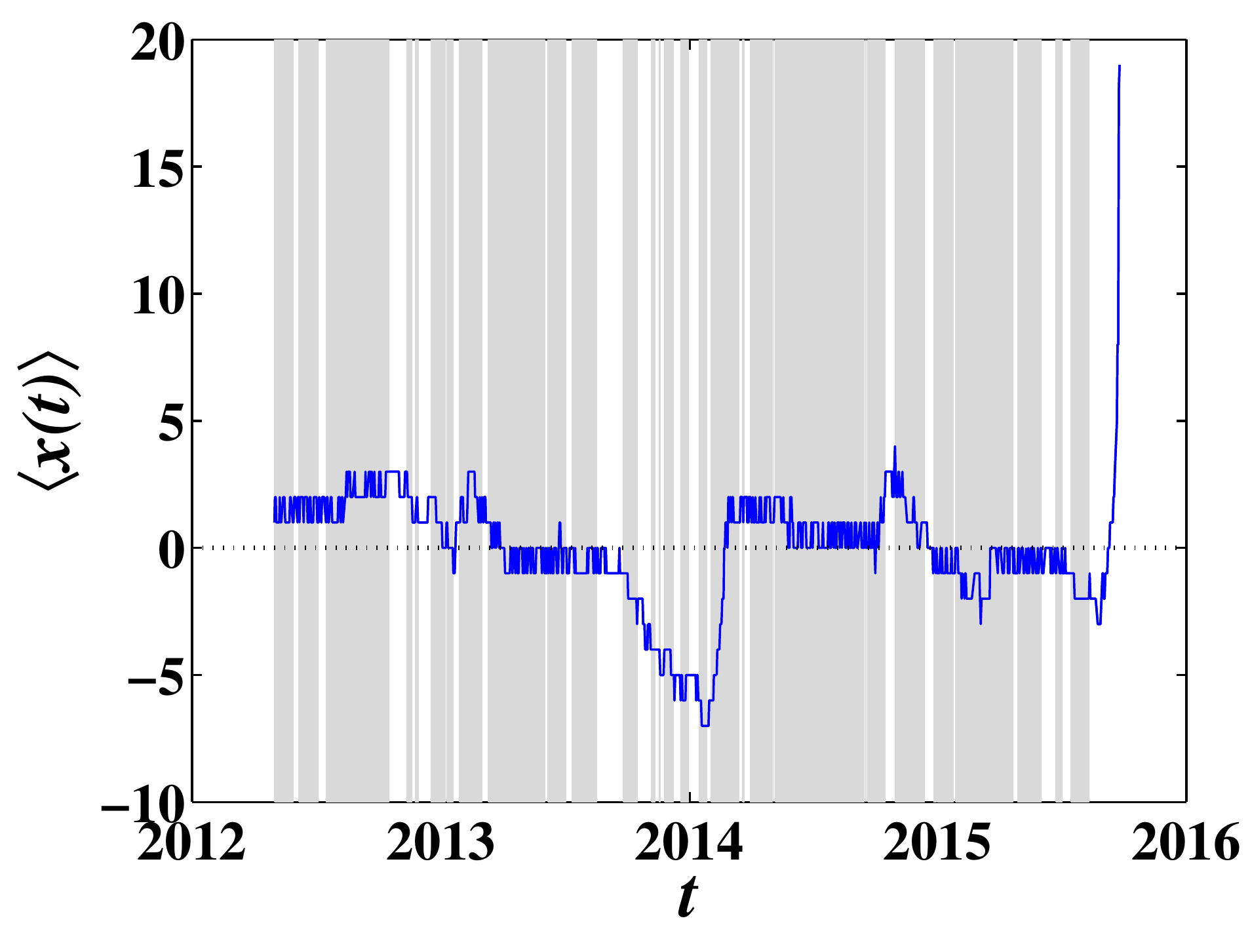}
  \includegraphics[width=0.32\linewidth]{Fig_Top_D_X_20_2.eps}\\
  \includegraphics[width=0.32\linewidth]{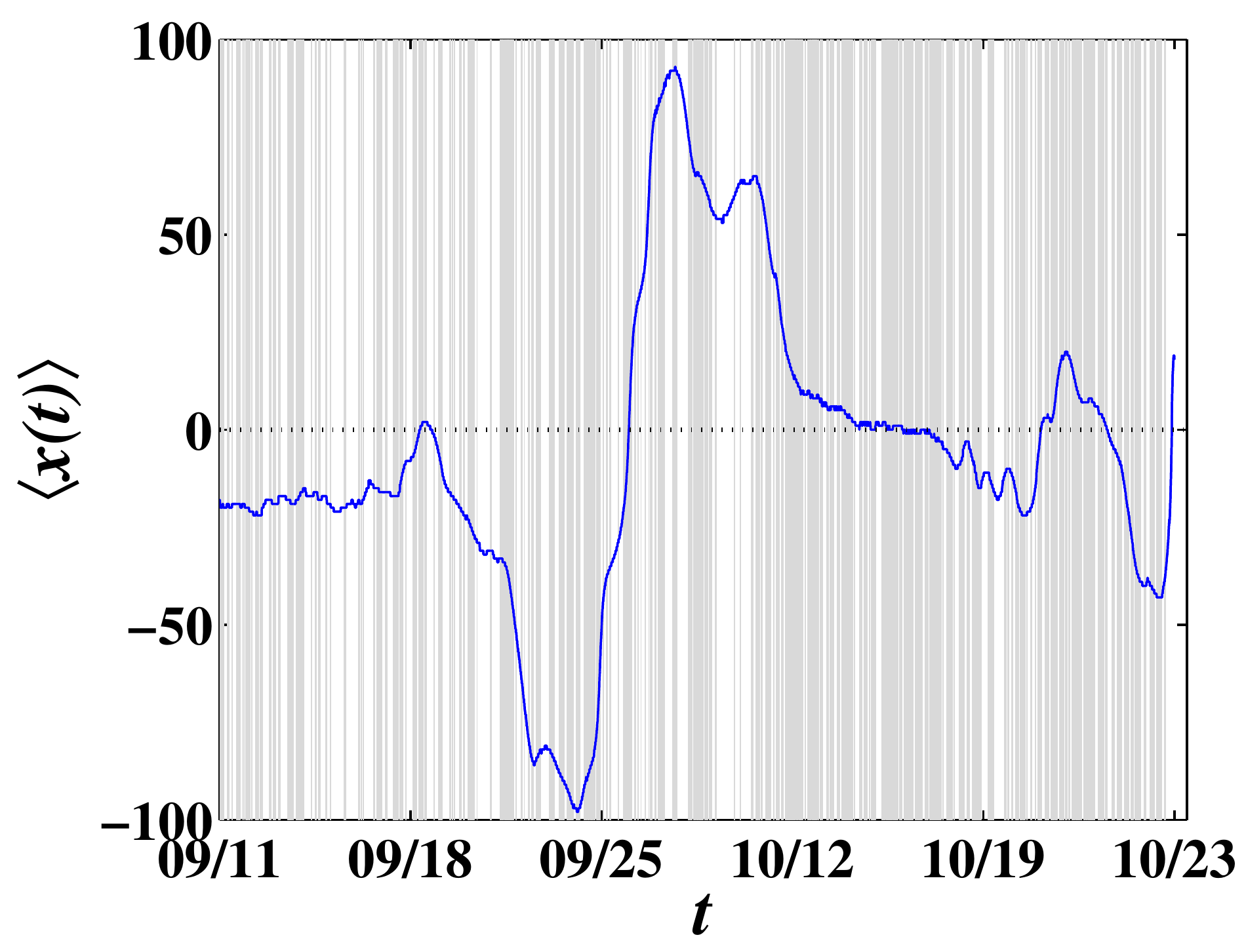}
  \includegraphics[width=0.32\linewidth]{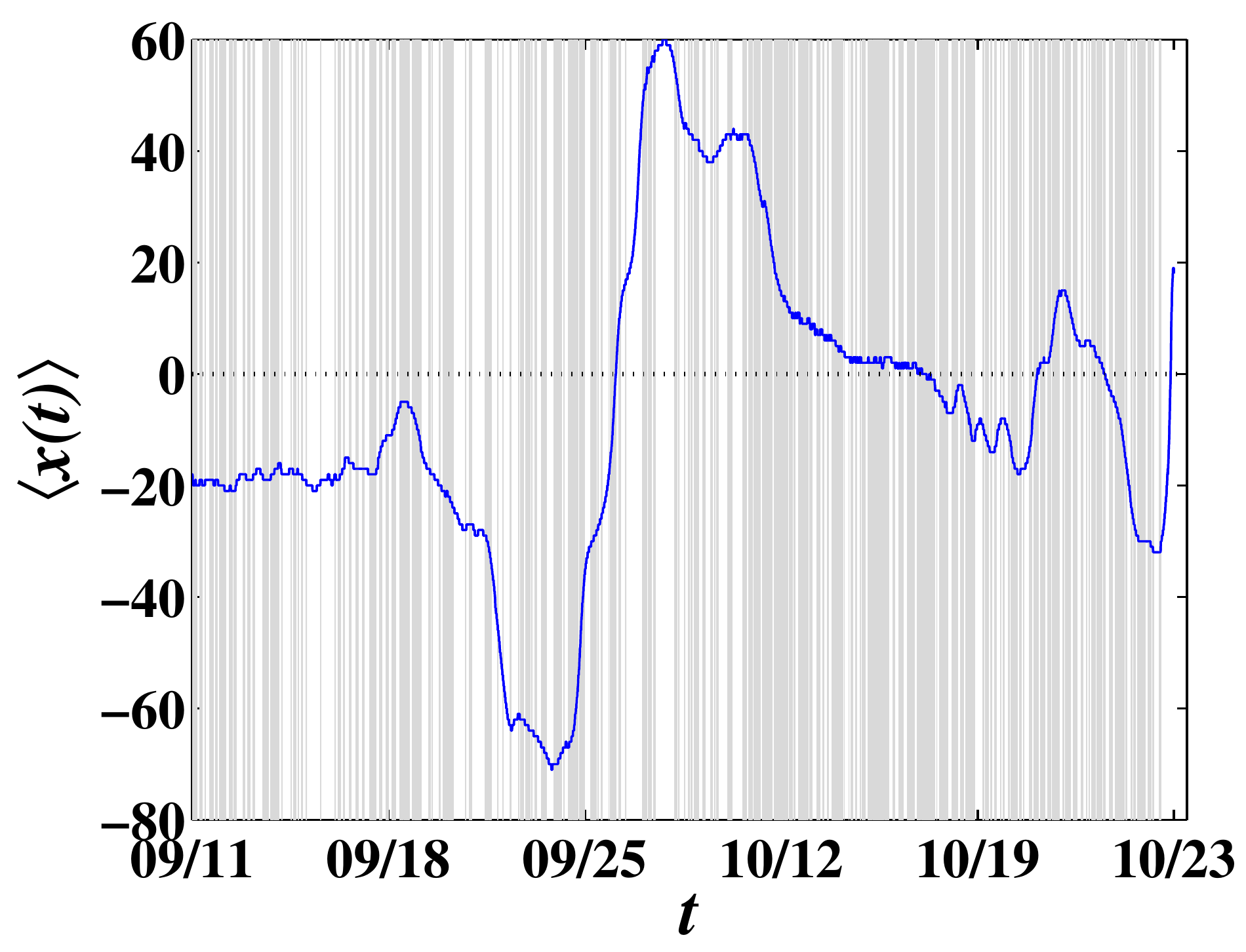}
  \includegraphics[width=0.32\linewidth]{Fig_Top_1m_X_20_2.eps}
  \vskip  -0.490\textwidth    \hskip   -0.97\textwidth (a)
  \vskip  -0.026\textwidth    \hskip  -0.3\textwidth (b)
  \vskip  -0.026\textwidth    \hskip  +0.36\textwidth (c)
  \vskip  +0.23\textwidth    \hskip   -0.97\textwidth (d)
  \vskip  -0.026\textwidth   \hskip   -0.3\textwidth (e)
  \vskip  -0.026\textwidth   \hskip   +0.36\textwidth (f)
  \vskip  +0.19\textwidth
  \caption{TOP analysis of the CNY and CNH exchange rates with two alternative values of parameter $T$. The grey shaded regions correspond to the time periods during which the consistency test is significant at the 5\% level. (a) Daily data and $T=1$. (b) Daily data and $T=1.5$. (c) Daily data and $T=2$. (d) Minute-scale data and $T=1$. (e) Minute-scale data and $T=1.5$. (f) Minute-scale data and $T=2$.}
  \label{Fig:TOP:T:1:1.5}
\end{figure}

Considering that the ``temperature'' $T$ is a key parameter of the TOP method that allows one to smooth out the
effect of noise and obtain more robust lead-lag relationship, it is important to test for the dependence of our results
on its chosen values. As mentioned, we allow ``thermal'' excitations or fluctuations around the absolute minimum energy path, so that path configurations with slightly larger global energies are allowed with probabilities decreasing with their energy. We specify the probability of a given path configuration with energy $\Delta E$ above the absolute minimum energy path by a so-called Boltzmann weight proportional to $\exp (-\Delta E /T)$, where $T$ quantifies how much deviations from the minimum energy are allowed. For $T \rightarrow 0$, the probability for selecting a path configuration of incremental energy $\Delta E$ above the absolute minimum energy path goes to zero. Increasing $T$ allows to sample more and more paths around the minimum energy path. Increasing $T$ thus allows us to wash out possible idiosyncratic dependencies of the path conformation on the specific realizations of the noises decorating the two time series. Of course, for too large temperatures, the energy landscape or distance matrix becomes irrelevant and one loses all information in the lag-lead relationship between the two time series. There is thus a compromise between not extracting too much from the spurious noise (a consideration that
pushes towards higher $T$) and washing out too much the relevant signal (which pushes towards lower $T$). \cite{Zhou-Sornette-2006-JMe} showed that, at temperatures $T>5$, the TOP method fails in general to extract the correct $\langle x(t) \rangle$. In this section, given other settings unchanged, $T$ is changed to two alternative values of 1 and 1.5. We present in Fig.~\ref{Fig:TOP:T:1:1.5} the results of the TOP analysis for daily and minute-scale CNY-CNH exchange rate with these two alternative $T$ values, as well as the results of $T=2$ in Fig.~\ref{Fig:TOP:Daily:2:20}(a) and Fig.~\ref{Fig:TOP:Minutely:2:20:100}(a). We observe that, although the TOP paths are not identical, they share the same lead-lag structure, recovering the main conclusions at the qualitative level.

For quantitative comparison, we consider how the ``temperature'' $T$ affect the lead-lag structure. Theoretically, we can infer that, with the increase of $T$, the lead-lag structure may be weakened since too many contributing paths are sampled. For daily exchange rates in Fig.~\ref{Fig:TOP:T:1:1.5}(a-c), the average optimal thermal paths $\langle x(t) \rangle$ show no obvious differences for different $T$. This is because the daily lead-lag structures are weak originally. However, for minute-scale exchange rates in Fig.~\ref{Fig:TOP:T:1:1.5}(d-f), we can observe that, the most positive $\langle x(t) \rangle$ becomes smaller with the increase of $T$ (93 for T=1, 60 for T=1.5, 42 for T=2), and the most negative $\langle x(t) \rangle$ also drops less with the increase of $T$ (-98 for $T=1$, -71 for $T=1.5$, -58 for $T=2$). This reflects the weakening effect of ``temperature'' $T$ on the lead-lag structure.

\begin{figure}[!htb]
  \centering
  \includegraphics[width=0.33\linewidth]{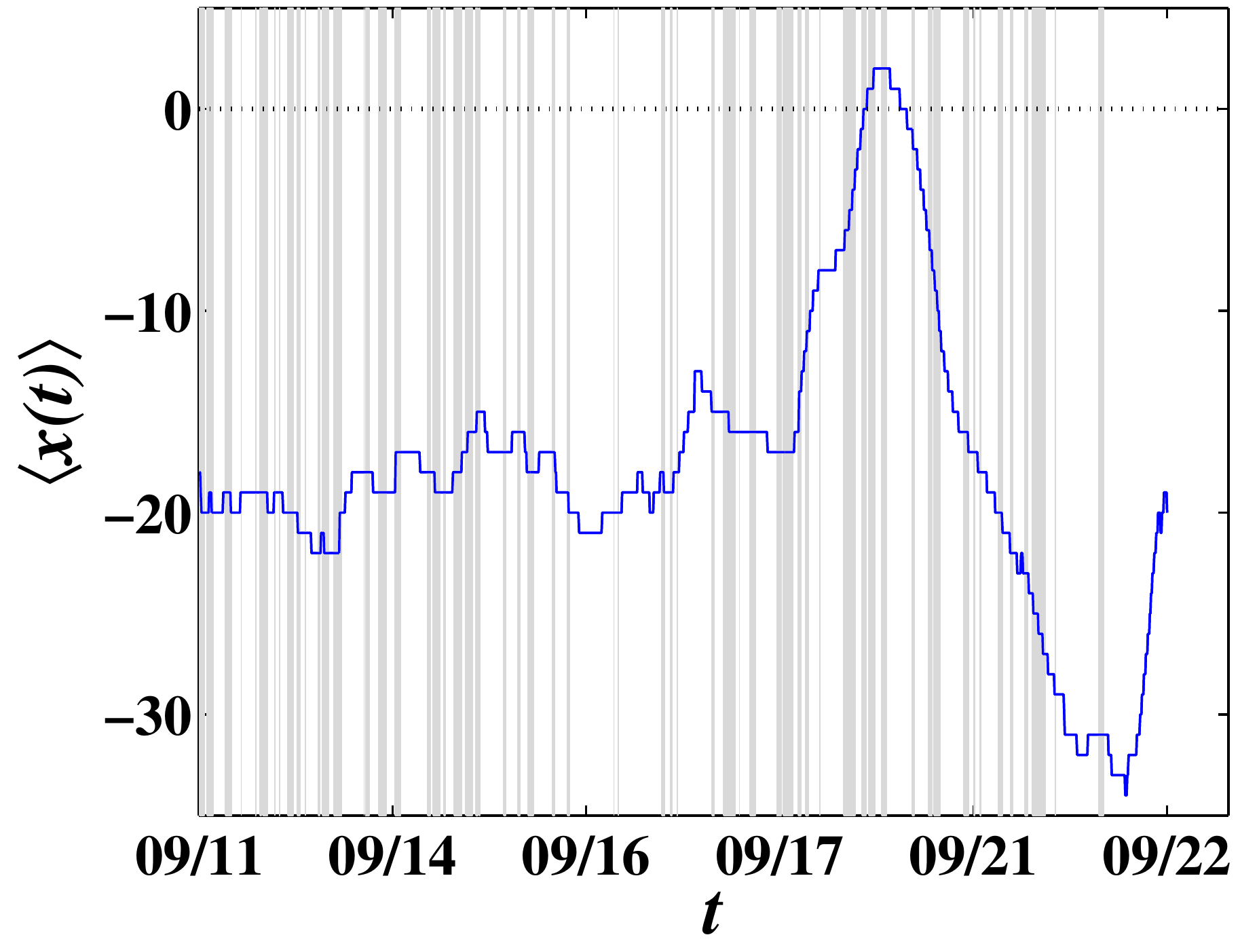}
  \includegraphics[width=0.33\linewidth]{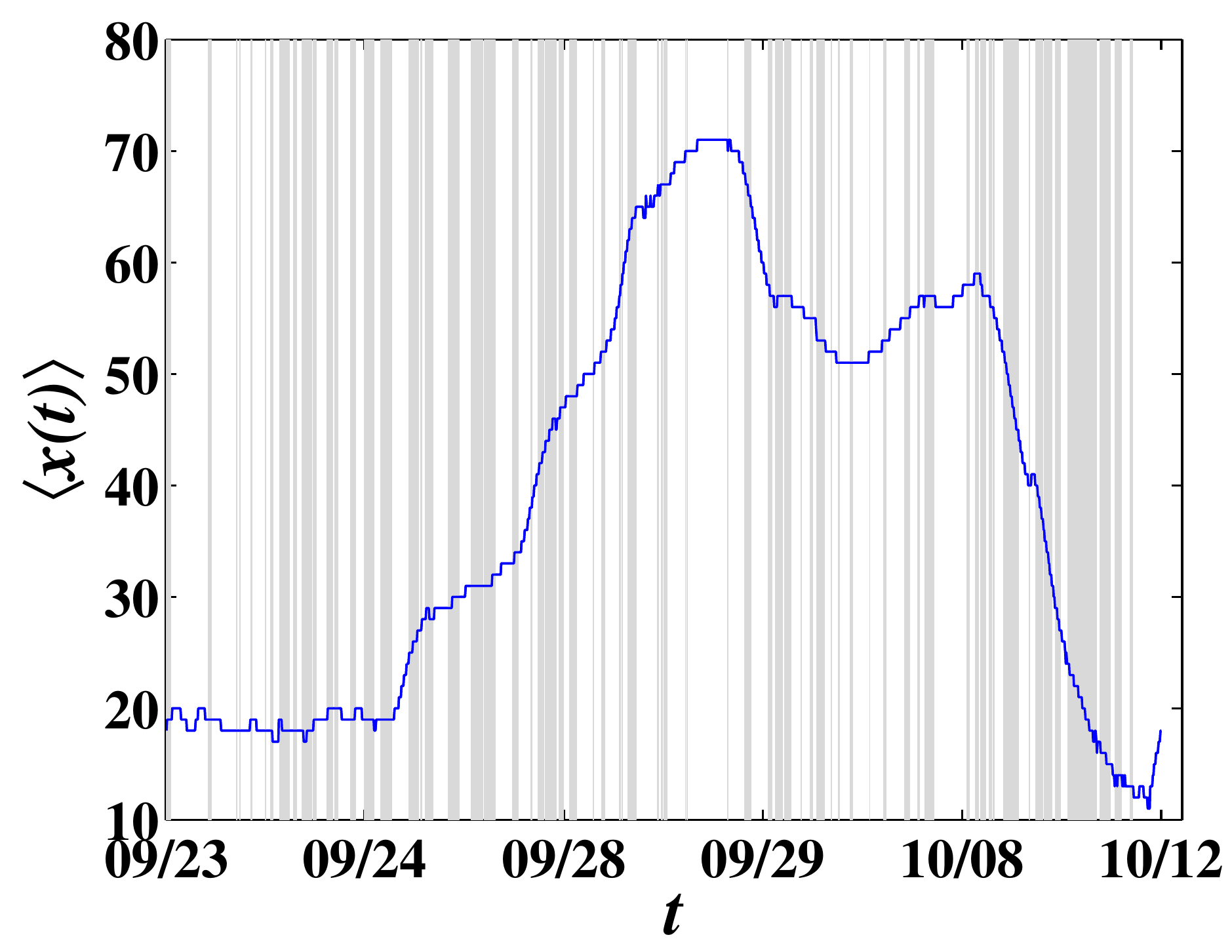}
  \includegraphics[width=0.33\linewidth]{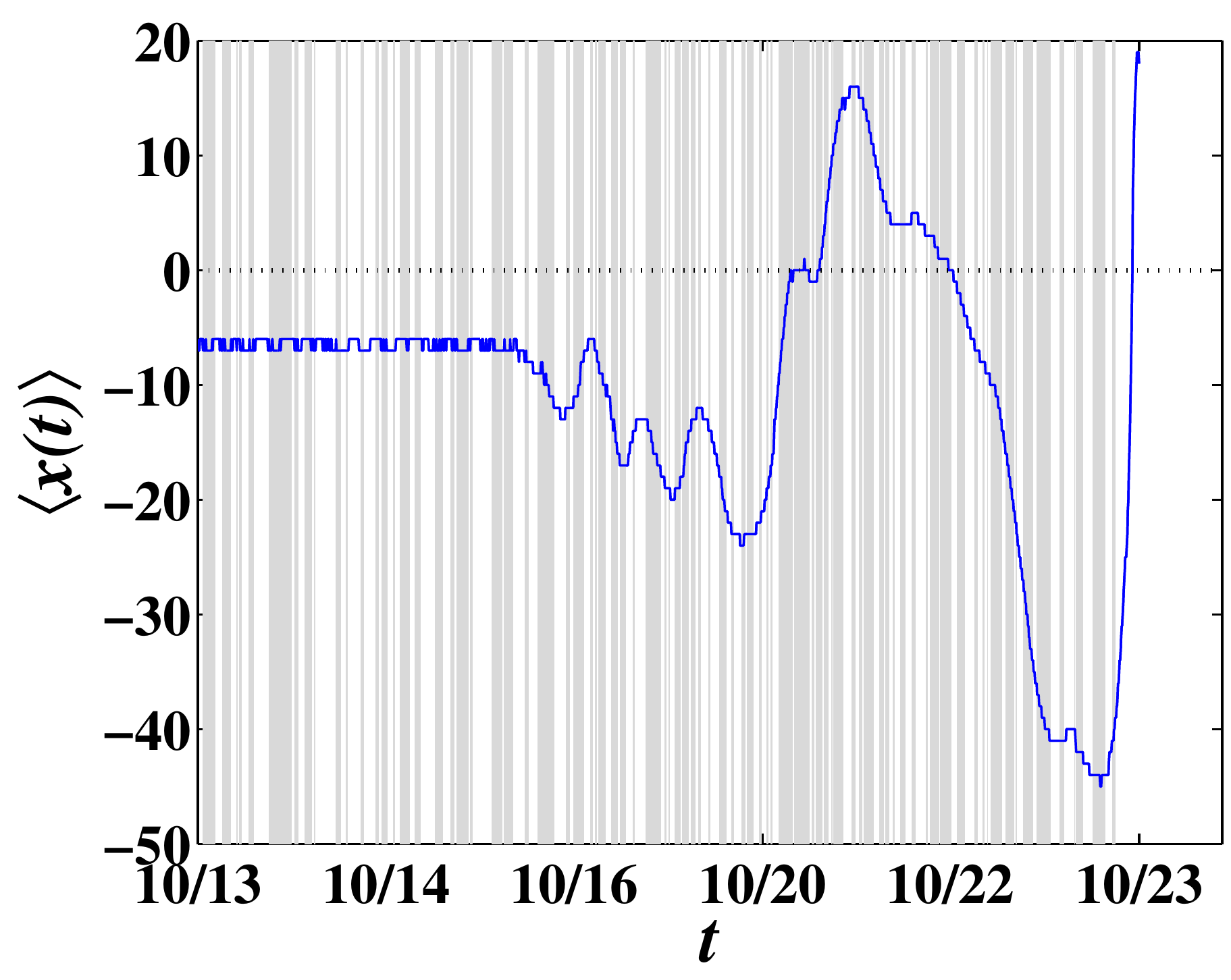}\\
  \includegraphics[width=0.33\linewidth]{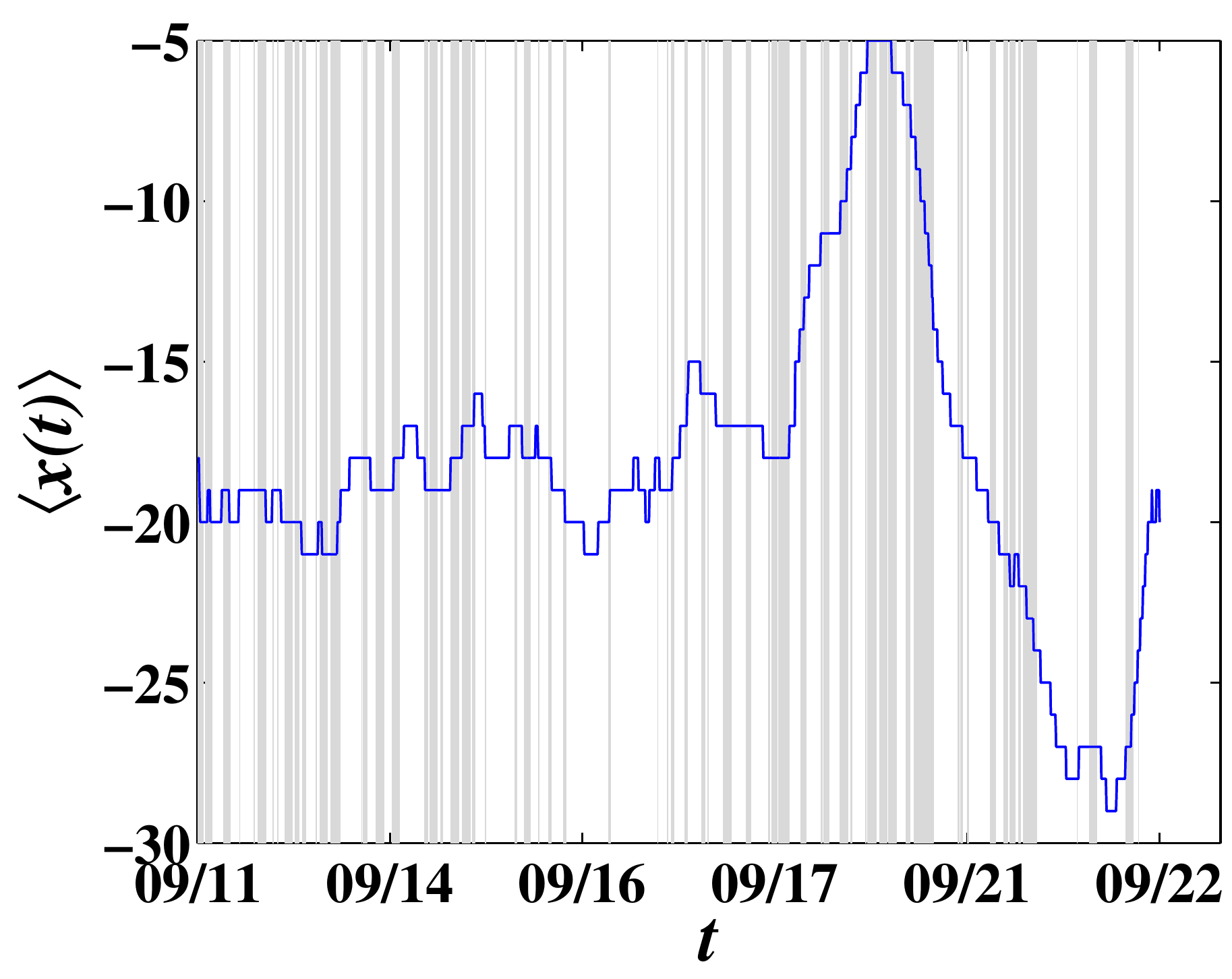}
  \includegraphics[width=0.33\linewidth]{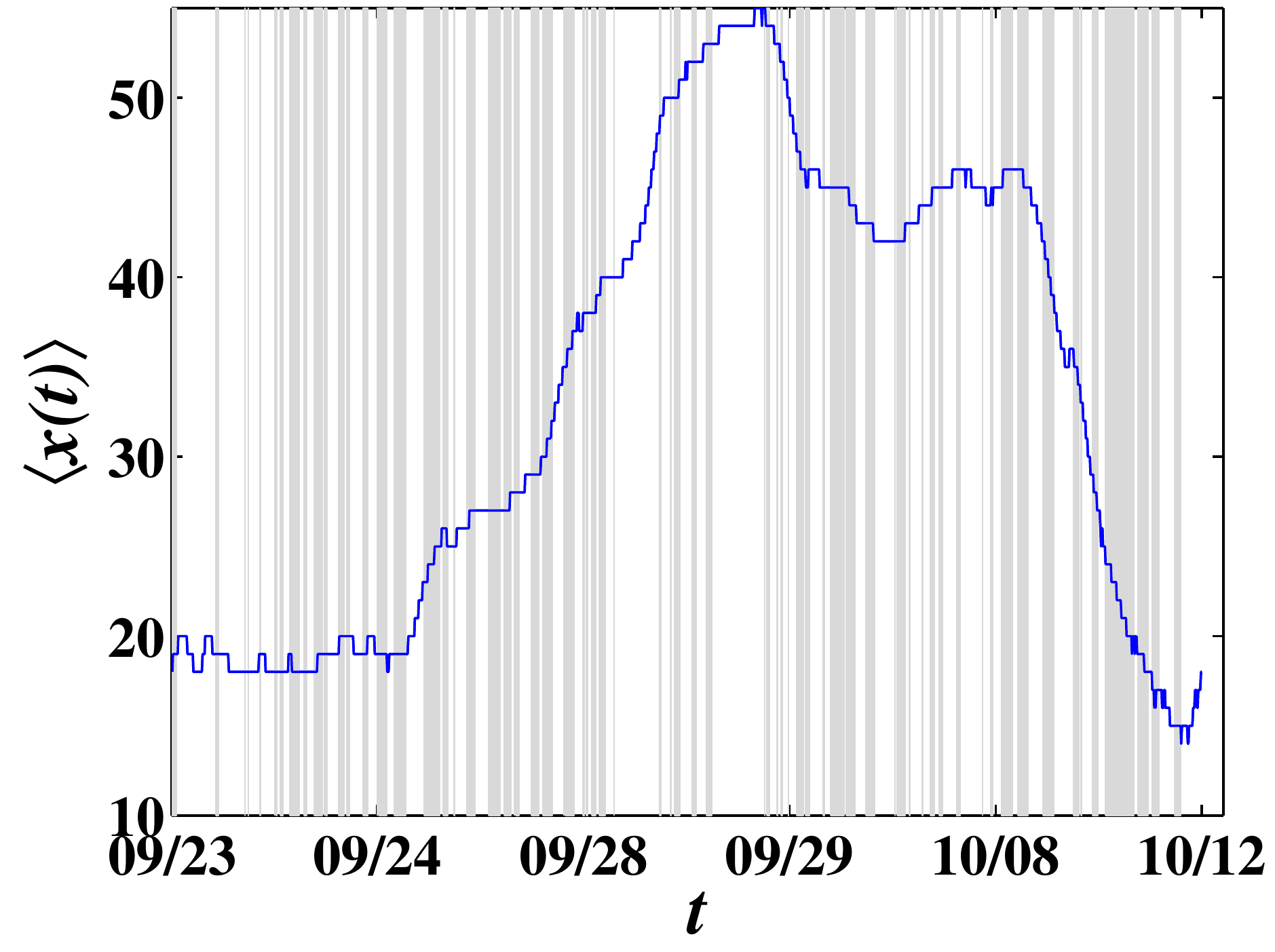}
  \includegraphics[width=0.33\linewidth]{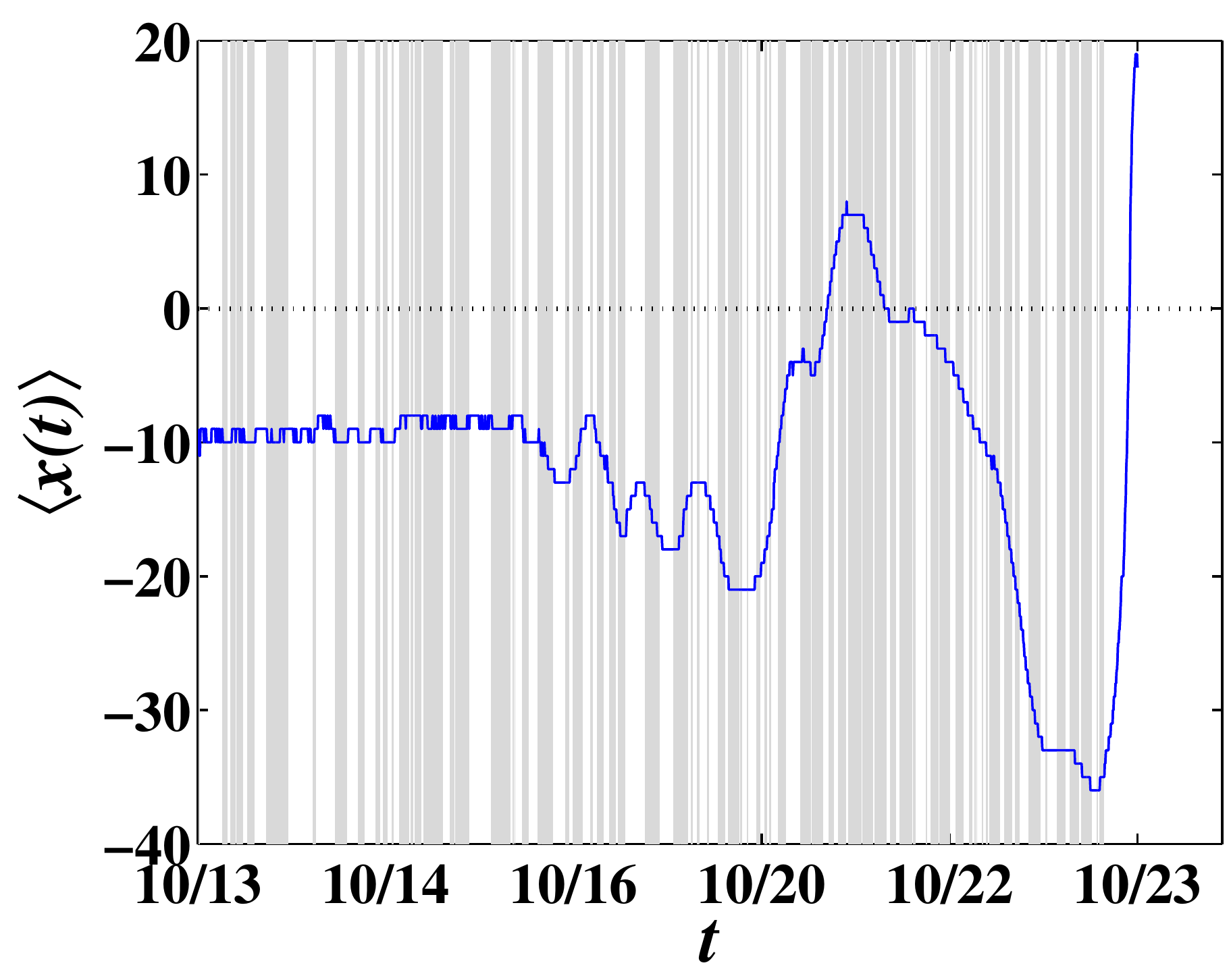}
  \vskip  -0.52\textwidth    \hskip   -0.96\textwidth (a)
  \vskip  -0.026\textwidth    \hskip  -0.3\textwidth (b)
  \vskip  -0.026\textwidth    \hskip  +0.36\textwidth (c)
  \vskip  +0.23\textwidth    \hskip   -0.96\textwidth (d)
  \vskip  -0.026\textwidth   \hskip   -0.3\textwidth (e)
  \vskip  -0.026\textwidth   \hskip   +0.36\textwidth (f)
  \vskip  +0.22\textwidth
  \caption{Sub-sample TOP analysis of the minute-scale CNY and CNH exchange rates with alternative values of parameter $T$. The grey shaded regions correspond to the time periods during which the consistency test is significant at the 5\% level. From left to right,
 the same three sub-periods as in Figure \ref{Fig:TOP:Sub:2:20} are shown.
 The top (resp. bottom) panels show the average optimal thermal path $\langle x(t) \rangle$ for $T=1$ (resp. $T=1.5$).}
  \label{Fig:TOP:Sub:T:1:1.5}
\end{figure}

We also perform sub-sample TOP analysis with alternative $T$ values. The results are shown in Fig.~\ref{Fig:TOP:Sub:T:1:1.5}. We can observe that the average optimal thermal paths $\langle x(t) \rangle$ are fairly stable with different $T$ such that our results are almost not influenced by $T$. This suggests that the CNY-CNH interaction patterns found above are credible.

\subsection{Alternative datasets}

As noted in previous section, the closing time of onshore market have been extended to 23:30 since 4 January 2016. This means that  in each day the corresponding trading period of CNH and CNY markets is also extended. To answer the question whether this time extension will influence our results, we perform the TOP analysis on an extra minute-scale data from 27 June 2016 to 11 July 2016.

\begin{figure}[!htb]
  \centering
  \includegraphics[width=0.45\linewidth]{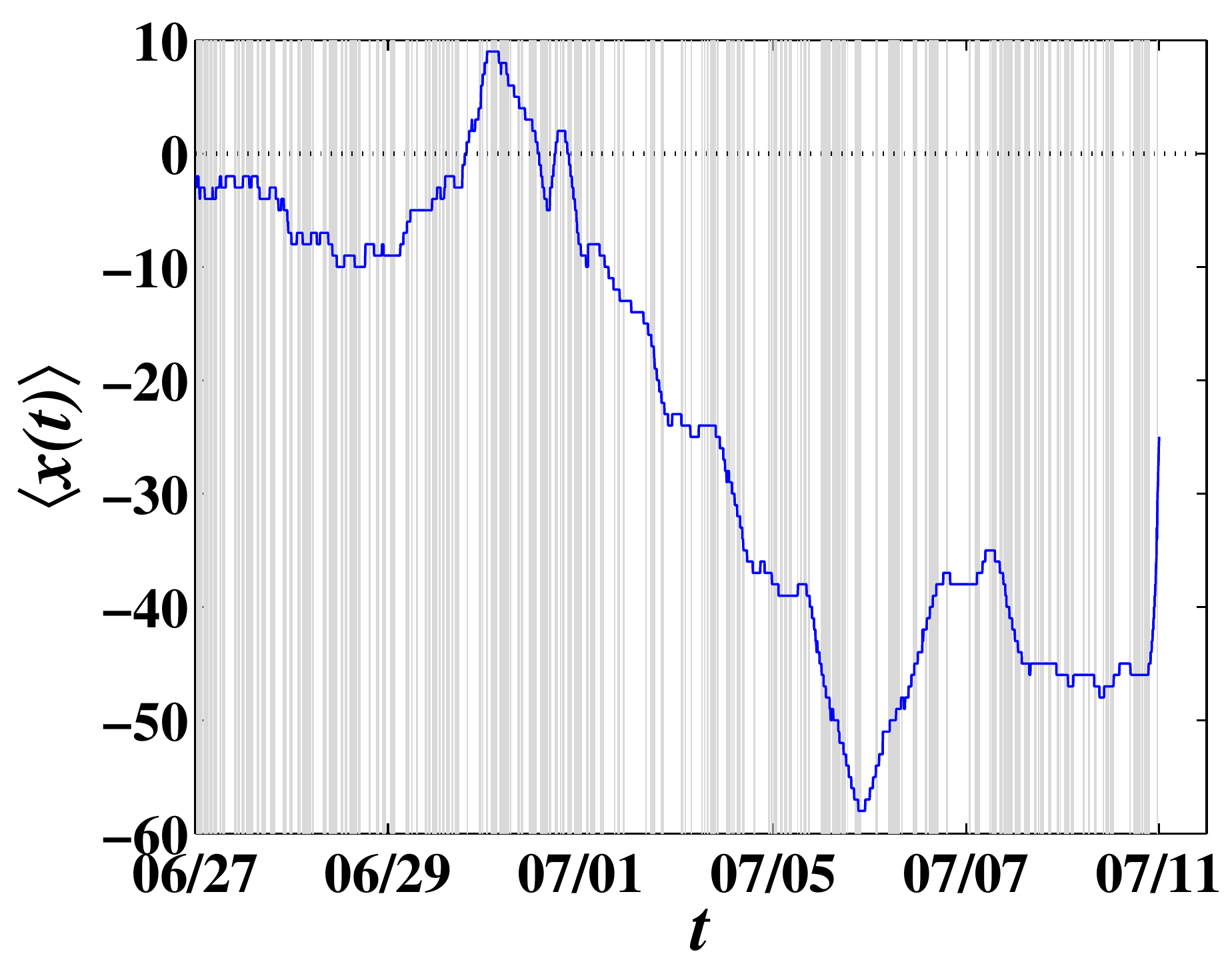}
  \includegraphics[width=0.45\linewidth]{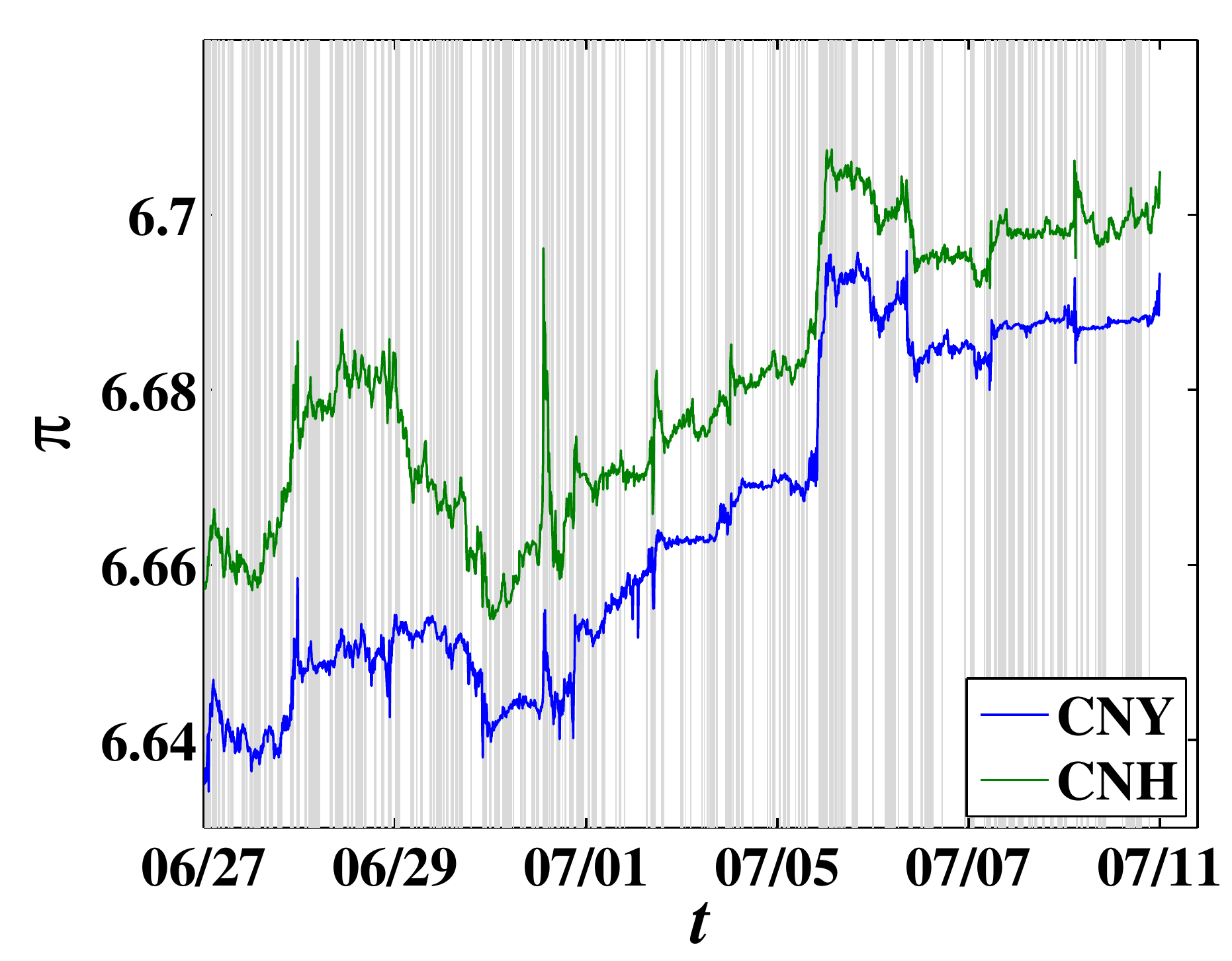}
  \caption{TOP analysis of the minute-scale CNY and CNH exchange rates from 27 June 2016 to 11 July 2016. The grey shades represent the periods when the consistency test is significant at the 5\% level. (a) Average optimal thermal path $\langle x(t) \rangle$ at temperature $T=2$. (b) CNY/USD (CNH/USD) exchange rates series $\pi$ at the one minute time scale as a function of time.}
  \label{Fig:TOP:Minite:SI:2:20}
\end{figure}

We present the TOP analysis for this supplementary data in Fig.~\ref{Fig:TOP:Minite:SI:2:20}. We observe that, generally the Renminbi is in a depreciation state (CNY/USD (CNH/USD) exchange rates $\pi$ rise up) during this period, and the majority of the average optimal thermal path $\langle x(t) \rangle$ are less than zero, indicating that the offshore (CNH) rates lead onshore (CNY). We should also notice that, there is a short period of Renminbi appreciation around 29-30 June 2016, and the counterparts of the average optimal thermal path $\langle x(t) \rangle$ turn from negative to positive, indicating that the onshore (CNY) rates lead offshore (CNH). These results are consistent with the analysis above. Therefore, the extension of the closing time will not change our results.

\section{Conclusions}
\label{S1:Conclusions}

Renminbi internationalisation has brought an active offshore market, which is closely related to the onshore market. Because the onshore market (CNY) is constrained by capital control and the existence of a trading band, the offshore exchange rate (CNH)  is generally believed to play the leading role in the price discovery of the onshore exchange market. However, this paper shows that this is not always correct.

We employ the thermal optimal path method to detect both the long-term (daily frequency) and short-term (minute-scale frequency) interaction patterns between the CNY and CNH exchange rates. At the daily level, most of the time, the CNY and CNH exchange rates show a weak alternating lead-lag structure. This reflects that fact that daily exchange rates of CNY and CNH track each other well. In several episodes in which CNY and CNH display a large disparity, the lead-lag relationship is uncertain and depends on the prevailing market factors.

The minute-scale interaction relationship between the CNY and CNH exchange rates also changes over time according to different short-term market situations. The lead or lag times are often up to tens of minutes. When the market is in stable or upward (Renminbi depreciation) periods, the average optimal thermal path is negative and thus the offshore exchange rate leads the onshore exchange rate. On the contrary, when the market is in downward (Renminbi appreciation) periods, the average optimal thermal path is positive and thus the onshore exchange rate leads the offshore exchange rate. As China's economy continues to slow down, Renminbi depreciation is a popular expectation of the markets. Therefore, investors may trade preferentially on the offshore market when
news occur, which explains the leading role of the offshore exchange rate when the market is in normal conditions. Conversely, the short-term appreciation of the Renminbi cannot be attributed to the offshore trading. This situation is more likely due to some macro factors or policy happening on the onshore market. Hence, under this situation, the onshore exchange rate plays a leading role over the offshore rate. Our results are partly consistent with the finding of \cite{Shu-He-Cheng-2015-CER}, who report that the CNH rate has greater impact on Asia-Pacific currencies in normal market conditions, but the CNY rate has greater influence during periods when the Renminbi appreciates against the US\$. Our findings on the interaction patterns between the CNY and CNH exchange rates are further confirmed by the sub-sample analysis and the alternative analysis with different smoothing parameters $T$. Note that for minute-scale rates, large $T$ weakens the detection of lead-lag structure.

Our study revealed both the long-term and short-term lead-lag relationship between the offshore and onshore exchange spot markets of the Chinese currency. For further analyses, it will be of interest to detect the time-dependent lead-lag relationships between the offshore forward markets and onshore spot markets, or relationships between the offshore options markets and onshore spot markets.

%


\bibliographystyle{elsarticle-harv}

\begin{thebibliography}{24}
\expandafter\ifx\csname natexlab\endcsname\relax\def\natexlab#1{#1}\fi
\expandafter\ifx\csname url\endcsname\relax
  \def\url#1{\texttt{#1}}\fi
\expandafter\ifx\csname urlprefix\endcsname\relax\def\urlprefix{URL }\fi

\bibitem[{Ashley et~al.(1980)Ashley, Granger, and
  Schmalensee}]{Ashley-Granger-Schmalensee-1980-Em}
Ashley, R., Granger, C. W.~J., Schmalensee, R., 1980. {Advertising and
  aggregate consumption - An analysis of causality}. Econometrica 48,
  1149--1167.

\bibitem[{{Bank for International Settlements}(2013)}]{BIS-2013-Basel}
{Bank for International Settlements}, 2013. {Triennial Central Bank Survey of
  Foreign Exchange and Derivatives Market Activity}. Basel.

\bibitem[{Cheung and Rime(2014)}]{Cheung-Rime-2014-JIMF}
Cheung, Y.-W., Rime, D., 2014. {The offshore renminbi exchange rate:
  Microstructure and links to the onshore market}. J. Int. Money Financ. 49,
  170--189.

\bibitem[{Chinn and Ito(2006)}]{Chinn-Ito-2006-JDE}
Chinn, M., Ito, H., 2006. {What matters for financial development? Capital
  controls, institutions, and interactions}. J. Dev. Econ. 81~(1), 163--192.

\bibitem[{Craig et~al.(2013)Craig, Hua, Ng, and
  Yuen}]{Craig-Hua-Ng-Yuen-2013-IMF}
Craig, R., Hua, C., Ng, P., Yuen, R., 2013. {Development of the Renminbi Market
  in Hong Kong SAR: Assessing Onshore-Offshore Market Integration}. IMF Working
  Paper.

\bibitem[{Derrida and Vannimenus(1983)}]{Derrida-Vannimenus-1983-PRB}
Derrida, B., Vannimenus, J., 1983. {Interface energy in random systems}. Phys.
  Rev. B 27, 4401--4411.

\bibitem[{Derrida et~al.(1978)Derrida, Vannimenus, and
  Pomeau}]{Derrida-Vannimenus-Pomeau-1978-JPC}
Derrida, B., Vannimenus, J., Pomeau, Y., 1978. {Simple frustrated systems:
  Chains, strips and squares}. J. Phys. C 11, 4749--4765.

\bibitem[{Ding et~al.(2014)Ding, Tse, and
  Williams}]{Ding-Tse-Williams-2014-JFutM}
Ding, D., Tse, Y., Williams, M., 2014. {The price discovery puzzle in offshore
  Yuan trading: different contributions for different contracts}. J. Fut.
  Markets 34~(2), 103--123.

\bibitem[{Funke et~al.(2015)Funke, Shu, Cheng, and
  Eraslan}]{Funke-Shu-Cheng-Eraslan-2015-JIMF}
Funke, M., Shu, C., Cheng, X., Eraslan, S., 2015. {Assessing the CNH-CNY
  pricing differential: role of fundamentals, contagion and policy}. J. Int.
  Money Financ. 59, 245--262.

\bibitem[{Geweke(1984)}]{Geweke-1984}
Geweke, J., 1984. Inference and causality in economic time series models. In:
  Griliches, Z., Intriligator, M. (Eds.), Handbook of Economics. Vol.~II.
  Elsevier Science Publisher BV, Amsterdam, Ch.~19, pp. 1101--1144.

\bibitem[{Gong et~al.(2016)Gong, Ji, Su, Li, and
  Ren}]{Gong-Ji-Su-Li-Ren-2016-PA}
Gong, C.-C., Ji, S.-D., Su, L.-L., Li, S.-P., Ren, F., 2016. {The lead-lag
  relationship between stock index and stock index futures: A thermal optimal
  path method}. Physica A 444, 63--72.

\bibitem[{Guo et~al.(2012)Guo, Zhou, and Cheng}]{Guo-Zhou-Cheng-2012-cnJMSC}
Guo, K., Zhou, W.-X., Cheng, S.-W., 2012. {Economy barometer analysis of China
  stock market: A dynamic analysis based on the thermal optimal path method}.
  J. Manag. Sci. China (in Chinese) 15~(1), 1--10.

\bibitem[{Guo et~al.(2011)Guo, Zhou, Cheng, and
  Sornette}]{Guo-Zhou-Cheng-Sornette-2011-PLoS1}
Guo, K., Zhou, W.-X., Cheng, S.-W., Sornette, D., 2011. {The US stock market
  leads the Federal funds rate and Treasury bond yields}. PLoS ONE 6~(8),
  e22794.

\bibitem[{Halpin-Healy and Zhang(1995)}]{HalpinHealy-Zhang-1995-PR}
Halpin-Healy, T., Zhang, Y.-C., 1995. {Kinetic roughening phenomena, stochastic
  growth directed polymers and all that}. Phys. Rep. 254, 215--415.

\bibitem[{Leung and Fu(2014)}]{Leung-Fu-2014-HKIMR}
Leung, D., Fu, J., 2014. {Interactions between CNY and CNH money and forward
  exchange markets}. HKIMR Working Paper.

\bibitem[{Li et~al.(2012)Li, Hui, and Chung}]{Li-Hui-Chung-2012-HKIMR}
Li, K.-F., Hui, C.-H., Chung, T.-K., 2012. {Determinants and dynamics of price
  disparity in onshore and offshore renminbi forward exchange rate markets}.
  HKIMR Working Paper.

\bibitem[{Maziad and Kang(2012)}]{Maziad-Kang-2012-IMF}
Maziad, S., Kang, J., 2012. {RMB Internationalization: Onshore/Offshore Links}.
  IMF working paper.

\bibitem[{Meng et~al.(2017)Meng, Xu, Zhou, and
  Sornette}]{Meng-Xu-Zhou-Sornette-2017-QF}
Meng, H., Xu, H.-C., Zhou, W.-X., Sornette, D., 2017. {Symmetric thermal
  optimal path and time-dependent lead-lag relationship: Novel statistical
  tests and application to UK and US real-estate and monetary policies}. Quant.
  Financ. 17, in press.

\bibitem[{Shu et~al.(2015)Shu, He, and Cheng}]{Shu-He-Cheng-2015-CER}
Shu, C., He, D., Cheng, X., 2015. One currency, two markets: the renminbi's
  growing influence in asia-pacific. China Econ. Rev. 33, 163--178.

\bibitem[{Sornette and Zhou(2005)}]{Sornette-Zhou-2005-QF}
Sornette, D., Zhou, W.-X., 2005. {Non-parametric determination of real-time lag
  structure between two time series: The ``optimal thermal causal path''
  method}. Quant. Financ. 5, 577--591.

\bibitem[{Wang(2015)}]{Wang-2015-JBF}
Wang, Y., 2015. {Convertibility restriction in China's foreign exchange market
  and its impact on forward pricing}. J. Bank. Financ. 50, 616--631.

\bibitem[{Zhang et~al.(2013)Zhang, Chau, and
  Zhang}]{Zhang-Chau-Zhang-2013-IRFA}
Zhang, Z., Chau, F., Zhang, W., 2013. {Exchange rate determination and dynamics
  in China: A market microstructure analysis}. Int. Rev. Financ. Anal. 29,
  303--316.

\bibitem[{Zhou and Sornette(2006)}]{Zhou-Sornette-2006-JMe}
Zhou, W.-X., Sornette, D., 2006. {Non-parametric determination of real-time lag
  structure between two time series: The ``optimal thermal causal path'' method
  with application to economic data}. J. Macroecon. 28, 195--224.

\bibitem[{Zhou and Sornette(2007)}]{Zhou-Sornette-2007-PA}
Zhou, W.-X., Sornette, D., 2007. {Lead-lag cross-sectional structure and
  detection of correlated-anticorrelated regime shifts: Application to the
  volatilities of inflation and economic growth rates}. Physica A 380,
  287--296.

\end{thebibliography}

\end{document}